\documentclass[superscriptaddress,12pt, preprint, a4paper,nofootinbib]{revtex4}
\usepackage{amsmath, amssymb, slashed, epsf}
\usepackage[dvipdfmx]{graphicx}
\usepackage{graphicx}
\usepackage{cancel}
\usepackage{array}
\usepackage{color}

\begin{document}
\renewcommand{\thefootnote}{\fnsymbol{footnote}}
\preprint{UT-HET-097}
\preprint{ICRR-REPORT-649-2014-20}
\preprint{KU-PH-015}

	\title{New resonance scale and fingerprint identification \\in minimal composite Higgs models }
	\author{Shinya Kanemura}
	\affiliation{Department of Physics, University of Toyama,\\ 
	3190 Gofuku, Toyama, 930-8555, Japan}
	\author{Kunio Kaneta}
	\affiliation{ICRR, University of Tokyo,\\ 
	Kashiwa, Chiba 277-8582, Japan}
	\author{Naoki Machida}
	\affiliation{Department of Physics, University of Toyama,\\ 
	3190 Gofuku, Toyama, 930-8555, Japan}
	\author{Tetsuo Shindou}
	\affiliation{Division of Liberal-Arts, Kogakuin University, \\
	1-24-2 Nishi-Shinjuku, Tokyo, 163-8677, Japan}
	\begin{abstract}
		Composite Higgs models are an intriguing scenario in which the Higgs particle is identified as a pseudo 
		Nambu-Goldstone boson associated with spontaneous breaking of some global symmetry above the electroweak scale.
		They would predict new resonances at high energy scales, some of which can appear at multi-TeV scales.
		In such a case, analogies with pion physics in QCD that a sizable phase shift is predicted in pion-pion scattering 
		processes might help us to evaluate scales of the resonances.
		In this paper, we discuss two complementary approaches to investigate the compositeness scale in minimal composite 
		Higgs models.
		First, we discuss the bound on vector boson scattering from perturbative unitarity, and we evaluate the phase shift 
		of the scattering amplitude, assuming that the same fitting function can be applied as the case in the pion physics.
		We then obtain the relation between possible phase shifts and promising new resonance 
		scales.
		We also investigate the possibility to measure the phase shift at LHC and the future hadron colliders.
		Second, we classify deviations in Higgs coupling constants from the standard model predictions in various kinds of the minimal composite Higgs models.
		We then discuss a possibility to discriminate a specific minimal composite Higgs model from the other models with extended Higgs sectors by utilizing deviation patterns in the Higgs boson couplings by future precision measurements.
	\end{abstract}

	\setcounter{footnote}{0}
	\renewcommand{\thefootnote}{\arabic{footnote}}
	\maketitle

\section{Introduction}

It is a crucial question whether the essence of the Higgs boson in the standard model (SM) is an elementary particle or a composite state.
The answer of this question gives a deeper insight to the fundamental theory of particle physics beyond the SM.
As a candidate of a new paradigm, supersymmetry has been extensively studied, where Higgs bosons are elementary scalar particles.
So far, however, no supersymmetric particle has been found by experiments, and  low-energy supersymmetric standard models are now being in trouble.
As an alternative, the Higgs boson can be a composite state, which is made of more fundamental fields by a certain strong dynamics.
The prototype of such a composite scenario is the technicolor model.
However, it has turned out to be challenging to construct a consistent model of the technicolor with the current experimental data.
After the discovery of the Higgs boson, particle physics enters a new era, where the essence of the Higgs boson can be explored directly by the measurement of Higgs boson properties.

Recently, composite scenarios again attract a lot of attention. 
In particular, the model originally proposed by Georgi and Kaplan \cite{Kaplan:1983fs,Kaplan:1983sm,Georgi:1984ef,Georgi:1984af,Dugan:1984hq} is revisited as a realistic candidate of new physics beyond the SM.
In this model, the Higgs boson is a pseudo Nambu-Goldstone boson (pNGB) associated with spontaneous breakdown of a global symmetry.
The Higgs boson mass is generated at the one-loop level due to the quantum effect of gauge bosons and fermions after the shift symmetry is explicitly broken by the $SU(2)_L\times U(1)_Y$ gauge symmetry.
Consequently, the mass 125 GeV could be naturally explained in this scenario.
The number of the NGB, $n$, is given by that of the broken generators, $n={\rm dim}(G/H)$, where the global symmetry $G$ is spontaneously broken down to the subgroup $H$ at a higher scale $f$ than the vacuum expectation value (VEV) of the Higgs field $v$.


In the minimal setup of the composite Higgs model, the global symmetry is $SO(5)\times U(1)_X$, in which $SO(5)$ is spontaneously broken into $SO(4)$ by any dynamics at the scale $f$~\cite{Agashe:2004rs}.
Models in this class are called the {\it minimal composite Higgs models} (MCHMs).
In these models, four NGBs appear, which correspond to the four component fields in the Higgs doublet of the SM.
A choice of the other symmetries with larger ${\rm dim}(G/H)$ can predict an extended Higgs sector in the low energy effective theory~\cite{Bellazzini:2014yua}.

We already know the existence of such pNGBs in Nature, which are the pions with spontaneous breakdown of the chiral symmetry.
The pions have masses corresponding to the explicit breakdown of the chiral symmetry.
The pion physics, the effective theory of QCD, has been well established as the chiral perturbation theory (see, e.g., \cite{Hatsuda:1994pi,Ecker:1994gg,Pich:1995bw} and references therein), which is a cutoff theory below the scale $4\pi f_\pi$ with $f_\pi$ being the pion decay constant.
It is known that perturbative unitarity is violated in pion-pion scatterings above the cutoff in the effective theory.
However, above the mass of the rho meson, the unitarity is rescued by the contribution of the diagram of the rho meson mediation.
In this case, a sizable phase shift would be induced in partial wave amplitudes of the pion-pion scatterings around the rho meson mass~\cite{Dobado:1989qm,Nebreda:2011di}.

In the MCHMs, from the analogy with the pion physics, we expect that there should be vector resonances in the scatterings of pNGBs; i.e., those of the Higgs boson and the longitudinally polarized weak gauge bosons.
In this paper, based on this hypothesis, we discuss to test MCHMs at future collider experiments.
In particular, we try to give answers to the following two interrelated questions: 
\begin{itemize}
	\item[(i)] How can we probe the scale of new resonance at future collider experiments?
	\item[(ii)] How can we distinguish the MCHMs from the future precision measurements of the Higgs boson coupling constants?
\end{itemize}
A promising  approach to the first question is to directly search a new resonance in scattering channels of weak gauge bosons and Higgs bosons. 
For example, the $WZ$ resonant channel is searched at LHC~\cite{Aad:2014pha,Khachatryan:2014xja}, which has given an exclusion limit on the resonance scale up to $1.5$ TeV.
The limit will be improved in near future at the LHC Run II with the collision energy 13-14 TeV.
The direct detection of the new resonance with a broad width will give a strong evidence of composite Higgs scenarios.
Even if a clear resonance cannot be observed, the information of the resonance scale would be indirectly obtained by measuring the phase shift of the scattering amplitudes as indicated in Ref.~\cite{Murayama:2014yja}.
In this paper, we discuss the possibility of extracting the resonance scale by measuring the phase shift in the scattering amplitude along with the method in Ref.~\cite{Murayama:2014yja} in the MCHMs.
In the MCHMs, in addition to the physics of the resonance in gauge boson scatterings, the coupling constants of the discovered 125 GeV Higgs boson in general deviate from the SM predictions as the consequence of physics of compositeness.
Thus, we can fingerprint MCHMs by detecting a pattern of deviation in Higgs boson couplings via precision measurements at future collider experiments such as the International Linear Collider (ILC).
To this end, in this paper, we investigate deviations in a set of the Higgs boson couplings in various MCHMs.
We therefore perform a complementary study in order to identify a specific model of MCHMs at future experiments.

In section \ref{sec:review}, we give a brief review of the matter independent part of MCHMs.
In section \ref{sec:phas shift}, we discuss perturbative unitarity in the MCHM.
The unitarity violation in the vector boson scattering amplitudes may indicate a new resonant state, where a sizable phase shift can be predicted similarly to the pion-pion scatterings.
We study the possibility to observe the phase shift in the $WZ$ production process at future hadron colliders such as the LHC Run II, high luminosity LHC with 3000 fb$^{-1}$ and also far future higher energy hadron colliders with the energy of 100 TeV, et cetera~\cite{Brock:2014tja}.
In section \ref{sec:fingerprint}, we discuss a method to discriminate the MCHMs by utilizing deviation patterns of the Higgs boson coupling constants.
We give a comprehensive list of deviations of Higgs boson couplings, which are to be checked by the precision measurements at future $e^+e^-$ collider experiments such as the ILC~\cite{Baer:2013cma,Asner:2013psa}, the Compact Linear Collider (CLIC)~\cite{Accomando:2004sz} and Future Circular Colliders (FCCs)~\cite{TLEP}.
In section \ref{sec:discussion}, we discuss the complementarity of these two approaches, and also we discuss the prospect for discriminating MCHMs from the other new physics models with extended Higgs sectors by fingerprinting patterns of deviations in the Higgs boson couplings.
The conclusions are given in section \ref{sec:summary}.
The kinematics used in section \ref{sec:phas shift} is given in Appendix A, and the definition of the variations of the MCHMs is given in Appendix B.

\section{Minimal composite Higgs models}
\label{sec:review}
We give a short review of the MCHMs \cite{Contino:2010rs,Bellazzini:2014yua} for the completeness and make our notation clear.
The Nambu-Goldstone bosons (NGBs) of $SO(5)\times U(1)_X\to SO(4)\times U(1)_X$ are parametrised as
\begin{equation}
	\Sigma=\Sigma_0 e^{\Pi/f}\;, \quad
	\Sigma_0=(0,0,0,0,1)\;,\quad
	\Pi=-iT^{\hat{a}}h^{\hat{a}}\sqrt{2}\;,
\end{equation}
where $f$ is a scale parameter analogous to $f_\pi$ in QCD and 
$h^{\hat{a}} (\hat{a}=1$-$4)$ denote the NGBs which correspond to 
the broken generators $T^{\hat{a}}$. We rewrite $\Sigma$ as 
\begin{equation}
	\Sigma =\frac{\sin(h/f)}{h}(h^1,h^2,h^3,h^4, h\cot(h/f)),
	\label{eq:HiggsSigma}
\end{equation}
where $h=\sqrt{h^{\hat{a}}h^{\hat{a}}}$.
It should be noted that $SO(4)$ ($\simeq SU(2)_L\times SU(2)_R$) contains a custodial symmetry.
We take the third component $h^3$ as the physical Higgs field.

We consider that a portion of $SO(5)\times U(1)_X$ is gauged, which corresponds to $SU(2)_L\times U(1)_Y$ in $SO(4)\times U(1)_X$.
The hypercharge $Y$ is given by the linear combination of a part of $SU(2)_R$ and $U(1)_X$: $Y=T_R^3+X$ with $T_R^3$ and $X$ being the eigenvalue of $SU(2)_R$ and the charge of $U(1)_X$, respectively.
Consequently, the global symmetry is explicitly broken by the gauge coupling, so that the NGBs become pNGBs.

The relevant part of the $SO(5)\times U(1)_X$ invariant effective 
Lagrangian is given by 
\begin{equation}
	\mathcal{L}=
	\frac{1}{2}P_{\mu\nu}\left[
		\Pi^X_0(p)X^{\mu}X^{\nu}
		+\Pi_0^{}(p)\text{Tr}[A^{\mu}A^{\nu}]
		+\Pi_1^{}(p)\Sigma A^{\mu}A^{\nu}\Sigma^T
	\right]
	+\mathcal{L}_{\text{matter}}
	\;,
\end{equation}
where 
$A_{\mu}$ ($=T^aA_{\mu}^a+T^{\hat{a}}A^{\hat{a}}_\mu$) are the $SO(5)$ gauge fields with $T^a$ being generators of $SO(4)$\footnote{
	In fact, the symmetry $SO(5)\times U(1)_X$ is a global symmetry and 
	only the $SU(2)_L\times U(1)_Y$ part of the symmetry is gauged.
	However, usually a trick is used such that a full $SO(5)\times U(1)_X$ symmetry 
	is assumed to be gauged, in order to write the Lagrangian in 
	simple $SO(5)\times U(1)_X$ invariant form.
	Only the $SU(2)_L\times U(1)_Y$ part of the $SO(5)\times 
	U(1)_X$ gauge fields are physical gauge fields and the rest are fake.
	Similar trick is used in the matter sector.
}, 
$X_{\mu}$ is the $U(1)_X$ gauge field, 
$\Pi$'s are the form factors, and
$P_{\mu\nu}=\eta_{\mu\nu}-p_{\mu}p_{\nu}/p^2$.
The interaction terms between $\Sigma$ and 
the matter fermions, which are denoted by $\mathcal{L}_{\text{matter}}$ in the above equation, 
depend on how to embed the SM matter fermions to $SO(5)$ 
representations. It provides us variety of the MCHMs.
We shall classify the MCHMs according to matter representations in section \ref{sec:fingerprint}.
The $SU(2)_L\times U(1)_Y$ invariant Lagrangian for the gauge sector is 
written in terms of the Higgs field $h$ as
\begin{align}
	\mathcal{L}_{\text{eff}}^{\text{gauge}}
	=&\frac{1}{2}P^{\mu\nu}\biggl[
		\left(\Pi_0^X(p)+\Pi_0(p)+\frac{\sin^2(h/f)}{4}\Pi_1^{}(p)\right)B_{\mu}B_{\nu}\nonumber\\
		&\phantom{\frac{1}{2}P_{\mu\nu}[]}
		+\left(\Pi_0^X(p)+\frac{\sin^2(h/f)}{4}\Pi_1^{}(p)\right)W_{\mu}^{a_L}W_{\nu}^{a_L}\nonumber\\
		&\phantom{\frac{1}{2}P_{\mu\nu}[]}
		+2\sin^2(h/f)\Pi_1^{}(p)\hat H^{\dagger}T^{a_L}Y\hat{H}W_{\mu}^{a_L}B_{\nu}
	\biggr]\;,
	\label{eq:gaugeLag}
\end{align}
where $W_{\mu}^{a_L}$ and $B_{\mu}$ are the $SU(2)_L$ and $U(1)_Y$ gauge fields, respectively, and the generators of $SU(2)_L$, $\{T^{a_L}\}$, are a partial set of $\{T^a\}$. 
Here, $\hat{H}$ can be expressed as
\begin{equation}
\hat{H}=\frac{1}{h}\begin{pmatrix} h^1-ih^2\\ h^3-ih^4\end{pmatrix}\;.
\end{equation}

The Higgs potential in the effective $SU(2)_L\times U(1)_Y$ gauge theory 
is generated at the one-loop level as 
\begin{equation}
	V_{\text{eff}}=V_{\text{eff}}^{\text{gauge}}+V_{\text{eff}}^{\text{fermion}}\;,
	\label{eq:HiggsPotential0}
\end{equation}
where $V_{\text{eff}}^{\text{gauge}}$ denotes the contributions from the 
$SU(2)_L$ gauge boson loops and $V_{\text{eff}}^{\text{fermion}}$ denotes
those from the SM matter fermion loops.
The contribution of the gauge boson loop is calculated as 
\begin{equation}
	V_{\text{eff}}^{\text{gauge}}=\frac{9}{2}\int\frac{d^4p}{(2\pi)^4}\ln \left(\Pi_0(p)^{}
	+\frac{1}{4}\Pi_1^{}(p)\sin^2(h/f)\right)\;.
	\label{eq:vgauge}
\end{equation}
If $V_{\text{eff}}^{\text{fermion}}$ is switched off, $SU(2)_L\times U(1)_Y$ is conserved at the minimum of the potential.
Therefore, $V_{\text{eff}}^{\text{fermion}}$, which depends on the matter sector of the MCHMs, $\mathcal{L}_{\text{eff}}^{\text{matter}}$, plays an important role in building a realistic model.

The gauge symmetry $SU(2)_L\times U(1)_Y$ is then broken by the Coleman-Weinberg mechanism due to quantum effects of the gauge fields and also matter fields~\cite{Coleman:1973jx}.
The electroweak symmetry breaking vacuum corresponds to 
\begin{equation}
	\langle \Sigma\rangle = \left(0,0,\sqrt{\xi},0,\sqrt{1-\xi}\right)\;,
\end{equation}
where the compositeness parameter $\xi$ is defined by $\sqrt{\xi}=\sin(\langle h\rangle/f)$.

By expanding $h$ around the VEV as $h\to \langle h\rangle +\hat{h}$, the effective Lagrangian given in Eq.~(\ref{eq:gaugeLag}) 
leads to the interaction terms between the Higgs boson $\hat{h}$ and 
the weak bosons as 
\begin{align}
	\mathcal{L}_{\text{eff}}^{\text{gauge}}=&
	\frac{g^2v^2}{4}W_{\mu}^+W^{-\mu}+\frac{g^2v}{2}\sqrt{1-\xi}\hat{h}W^+_{\mu}W^{-\mu}
	+\frac{g^2(1-2\xi)}{4}\hat{h}^2W_{\mu}^+W^{-\mu}\nonumber\\
	&+\frac{g_Z^2v^2}{4}Z_{\mu}Z^{\mu}+\frac{g_Z^2v}{2}\sqrt{1-\xi}\hat{h}Z_{\mu}Z^{\mu}
	+\frac{g_Z^2(1-2\xi)}{4}\hat{h}^2Z_{\mu}Z^{\mu}\;,
	\label{eq:HiggsGaugeCoupling}
\end{align}
where $g_Z^2=g^2+g^{\prime 2}$ and $v=f\sqrt{\xi}$ $(= 246~\text{GeV})$.
The couplings between the Higgs boson and the gauge bosons are determined only by 
the breaking pattern of the symmetry as $SO(5)\times U(1)_X\to SO(4)\times 
U(1)_X$, and one can find that the Higgs couplings to the weak bosons deviate from the 
SM predictions as $g_{hVV}=g_{hVV}^{\text{SM}}\sqrt{1-\xi}$, which are independent of the matter sector. 
The size of the compositeness parameter $\xi$ is constrained by the LEP precision data.
As shown in Ref.~\cite{Contino:2010mh}, $\xi>0.2$ has been excluded in the 99\% confidence level in the typical case of the MCHM\footnote{Of course, the limit from the precision data depends on the matter sector, but here we omit the matter dependence which might make the limit weaker or stronger.}.

\section{New resonance scale and sizable phase from perturbative unitarity}
\label{sec:phas shift}

\begin{figure}[htbp]
\begin{center}
\hspace*{6mm}
\includegraphics[width=70mm]{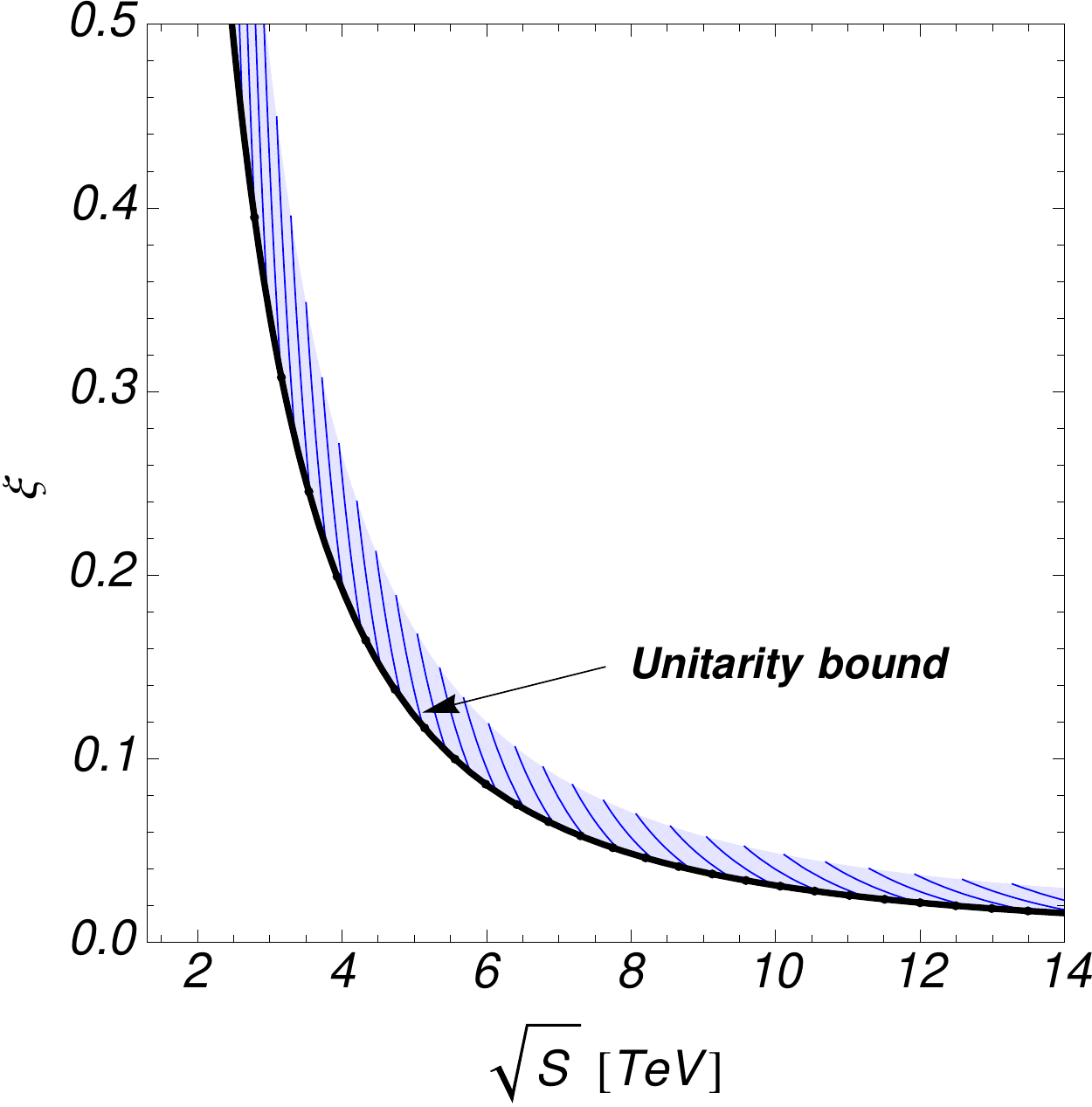}
\end{center}
\caption{The unitarity bound as a function of $\xi$ and $\sqrt{S}$ given in Eq.(\ref{eq:unitarity}). The region above the solid curve forbidden by violating perturbative unitarity.}
\label{fig:unitarity}
\end{figure}

In the MCHMs, as shown in section \ref{sec:review}, the Higgs boson couplings with the weak gauge bosons deviate from the SM predictions; i.e., $g_{hVV}=g_{hVV}^{\rm SM} \sqrt{1-\xi}$ and $g_{hhVV}=g_{hhVV}^{\rm SM}(1-2\xi)$.
Intriguingly, the deviations break perturbative unitarity of elastic scatterings of the longitudinal component of weak gauge bosons, $W_L$, at high energies.
The unitarity argument gives the bound on partial wave amplitudes $a_l$ in the channel of the angular momentum $l$:
\begin{eqnarray}
	{\rm Re}[a_l]^2+\left({\rm Im}[a_l]-\frac{1}{2}\right)^2= \left(\frac{1}{2}\right)^2, \label{eq:unitarity_0}
\end{eqnarray}
which gives a circle of the radius $1/2$ with the center at $(0,1/2)$ in the complex-plane of $a_l$, and thus the real part of the amplitude can not exceed $1/2$.
Let us consider the s-wave amplitude for the elastic scattering of $W_L$,
\begin{eqnarray}
	a_0 &=& \frac{G_F\xi S}{16\sqrt{2}\pi} + \frac{G_F(m_h^2 - M_W^2)(1-\xi)}{4\sqrt{2}\pi} ~~~~~(\sqrt{S}\gg m_h), \label{eq:unitarity}
\end{eqnarray}
where $\sqrt{S}$ is the center-of-mass energy of the scattering.
The unitarity bound is given by $\left| a_0\right| \leq \frac{1}{2}$.
In MCHMs, perturbative unitarity is violated by the non-vanishing compositeness parameter $\xi$. 
In Fig.~\ref{fig:unitarity}, the unitarity bound is shown on the $\sqrt{S}$--$\xi$ plane, where perturbative unitarity is violated in the region beyond the solid curve.
This limit tells us that the amplitude increases in proportion to $S$ due to the non-vanishing $\xi$, and goes over the unitarity bound above some scales.

It is of importance to explore the validity of the SM at high energies from the argument of unitarity in scattering amplitudes of longitudinally polarized weak gauge bosons~\cite{Contino:2010mh}.
If we observe that the tail of the $W_LW_L$ scattering cross section becomes larger than the SM prediction at high energies, there can be the following two possibilities: (i) the dynamics of electroweak symmetry breaking becomes a strongly coupled theory at high energies where perturbation calculation cannot be applied. (ii) appearance of a new resonance with a broad width at a little bit above the scale, which saves unitarity above the resonance\footnote{In the weakly coupled theory with UV completion, a resonance of new particles has a sharp peak with a narrow width. 
This case may be a similar situation to the Fermi theory where the cutoff scale is the $W$ boson mass.}.
In the case of the second possibility, we may be able to extract the information of the resonance by measuring the phase shift of the scattering amplitude.
The case (ii) corresponds to an analogy of the rho meson in a viewpoint of the effective theory of the pion.

In the rest of this section, we focus on the second case (ii), and discuss the energy scale of a new resonance from the phase shift of the scattering amplitude.
Due to the imaginary part of the scattering amplitude, a sizable phase shift appears in scattering processes as discussed in Refs.~\cite{Dobado:1989qm,Barklow:1997nf,Murayama:2014yja,Delgado:2014dxa}.
In general, the profile of the phase shift is unknown for the $W_LW_L$ scattering.
On the other hand, for the pion-pion elastic scattering, as shown in Ref.~\cite{Murayama:2014yja}, it is fitted by
\begin{eqnarray}
	\delta &=& \left\{
	\begin{array}{l}
		\tan^{-1}\displaystyle{\left[\frac{\Gamma}{m}\frac{S}{m^2+\Gamma^2-S}\right]} \qquad\quad{\rm for}\quad \sqrt{S}<\sqrt{m^2+\Gamma^2},\\[6mm] 
		\tan^{-1}\displaystyle{\left[\frac{\Gamma}{m}\frac{S}{m^2+\Gamma^2-S}\right]}+\pi \quad{\rm for}\quad \sqrt{S}\geq \sqrt{m^2+\Gamma^2},
	\end{array}
	\right.\label{phase1}
\end{eqnarray}
where $m$ and $\Gamma$ are the rho meson mass and its decay width, respectively.
From the similarity in physics of pNGBs between in the pion physics and in the MCHMs, in the following, we dare to assume that the same fitting function can be applied to the $W_LW_L$ scattering in the MCHMs as in Ref.~\cite{Murayama:2014yja}.

When the increasing behavior of the amplitude is supposed to be a tail of new resonance, the phase shift is given by the argument of the amplitude:
\begin{eqnarray}
	\delta
	&\equiv&\tan^{-1}\left[\frac{{\rm Im}[a_0]}{{\rm Re}[a_0]}\right]
	= \tan^{-1}\left[
	\frac{1}{2{\rm Re}[a_0]}\pm\sqrt{\frac{1}{4{\rm Re}[a_0]^2}-1}
	\right],\label{phase2}
\end{eqnarray}
where, by using Eq.~(\ref{eq:unitarity_0}), ${\rm Im}[a_0]$ can be expressed in terms of ${\rm Re}[a_0]$ which is given in Eq.~(\ref{eq:unitarity}).
It should be noted that Eq.~(\ref{phase2}) gives a typical value for $\delta$ since the amplitude $a_0$ is located on the unitarity circle given in Eq.~(\ref{eq:unitarity_0}) (not within the circle for elastic scatterings).
In other words, this condition gives a conservative limit for the allowed imaginary part of the amplitude since the amplitude becomes maximally large on the unitarity circle.

\begin{center}
\begin{figure}[htbp]
	\begin{tabular}{cc}
		\includegraphics[width=70mm]{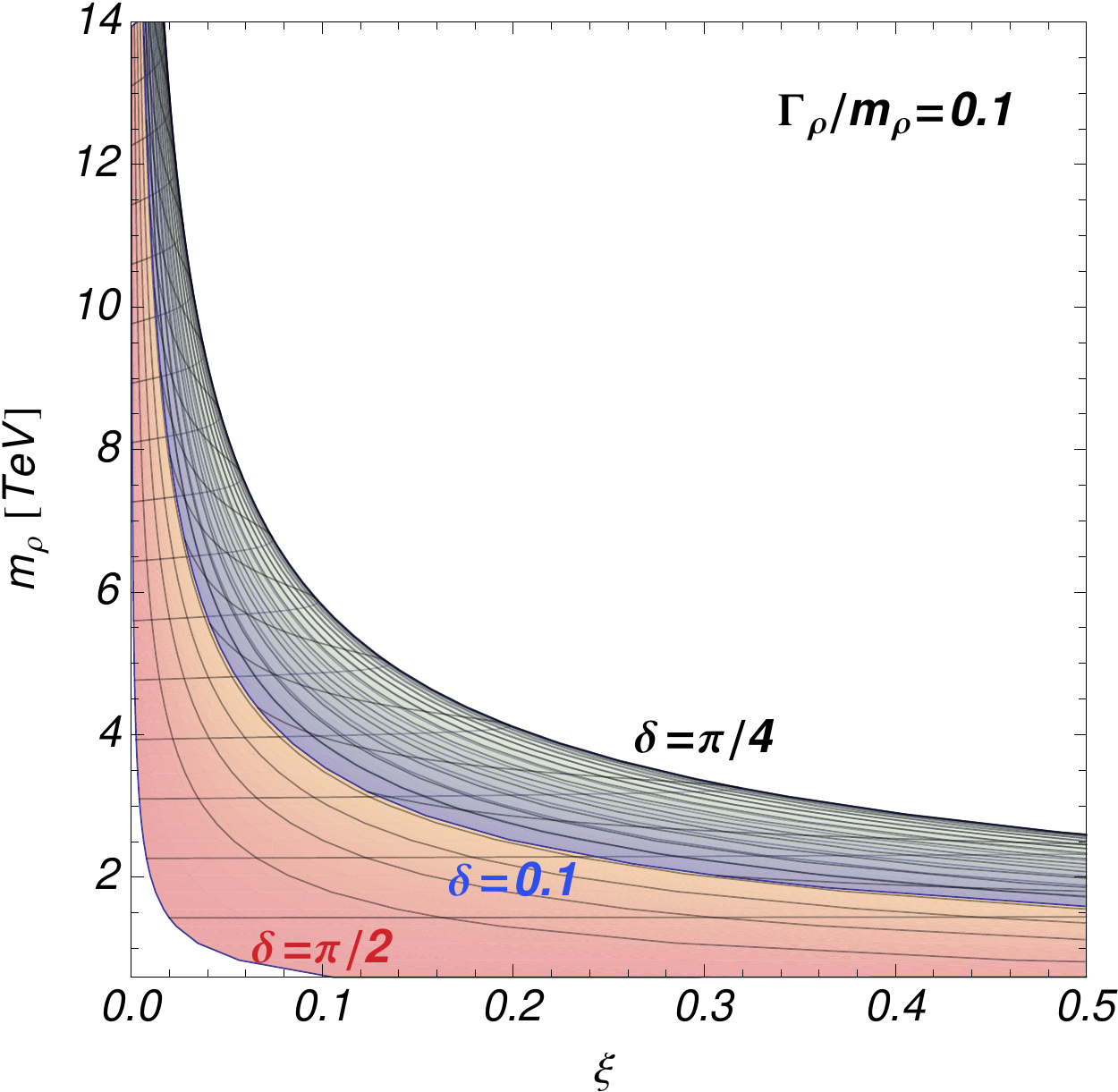}&
	\includegraphics[width=70mm]{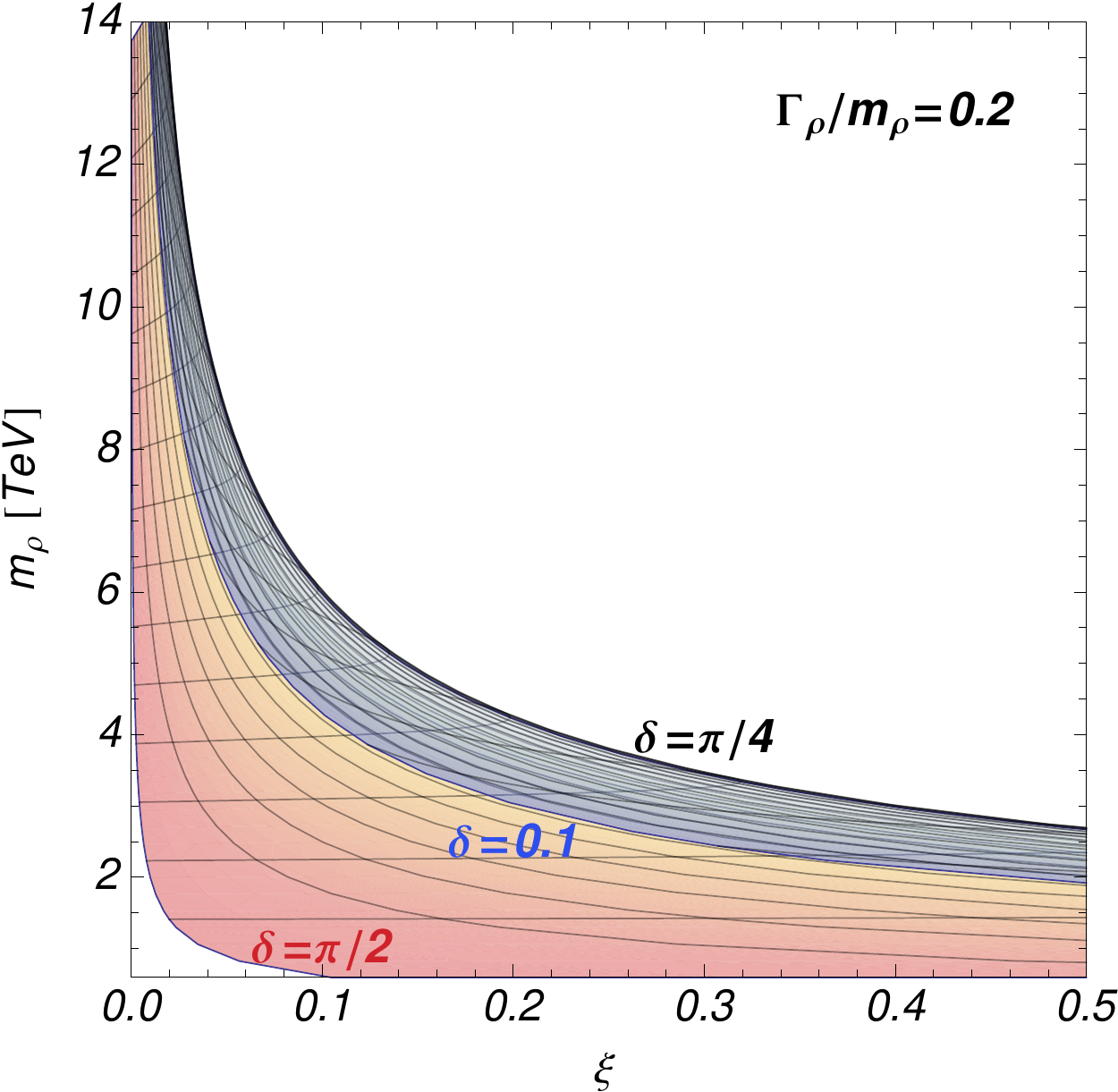}\\
	\\
	\includegraphics[width=70mm]{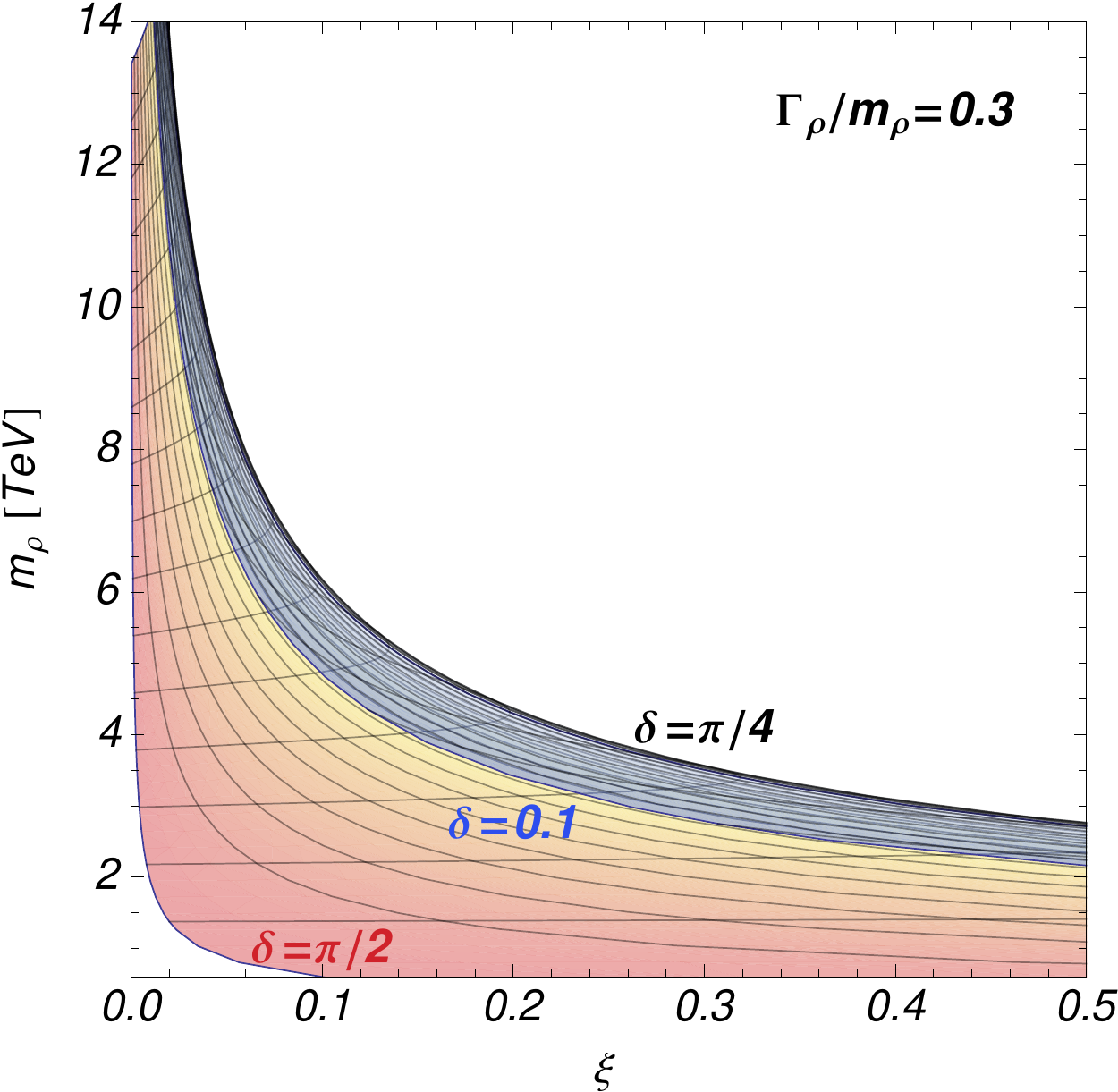}&
	\\
\end{tabular}
\caption{Three panels show new resonance scales predicted by perturbative unitarity, where a sizable phase shift appears.
We take the width-to-mass ratio as $\Gamma_\rho/m_\rho=0.1,0.2$ and $0.3$ in the upper-left, upper-right and bottom-left panels, respectively.}
\label{fig:phase}
\end{figure}
\end{center}

\begin{center}
\begin{figure}[htbp]
\includegraphics[width=95mm]{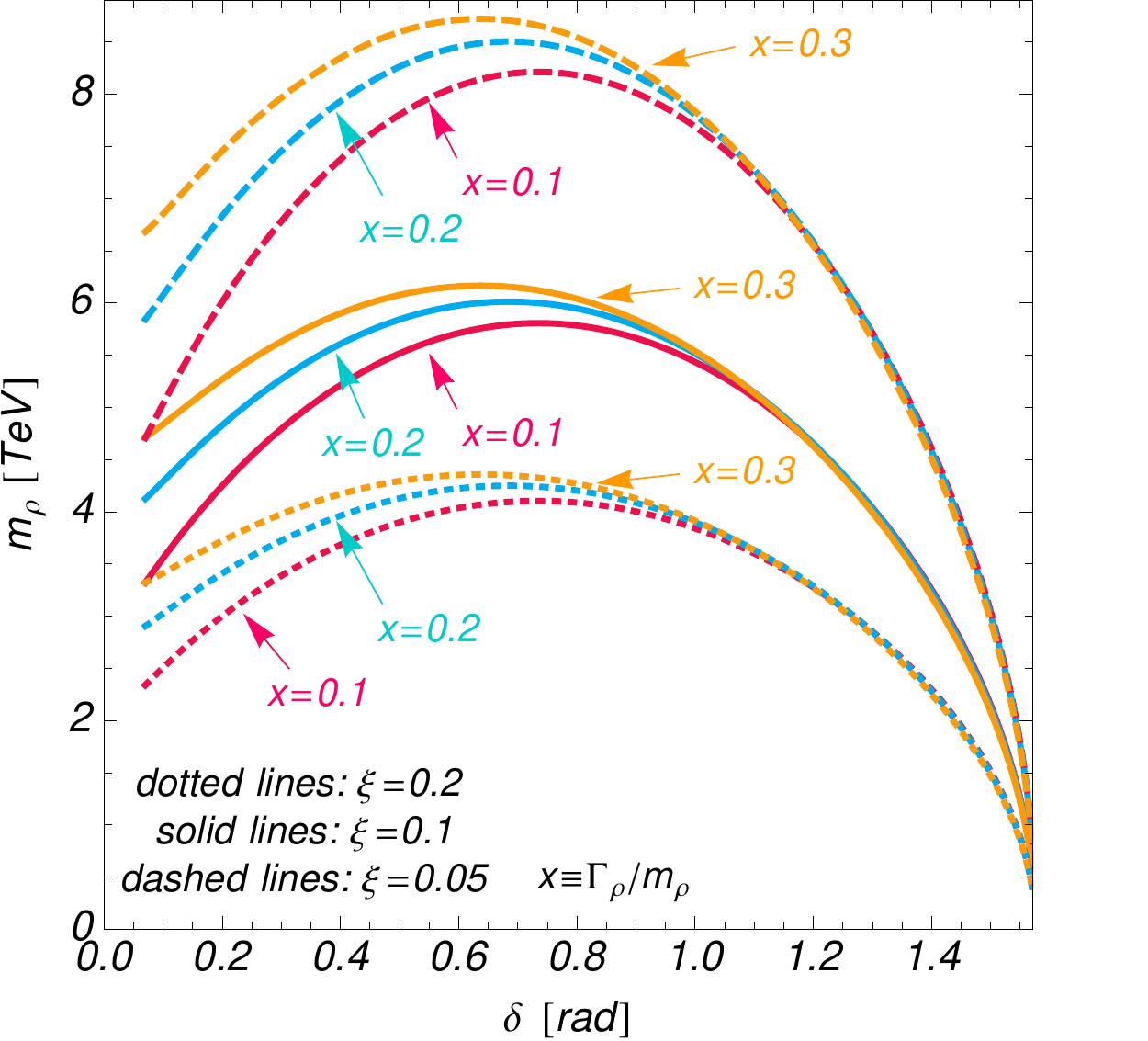}
\caption{The relation between $m_\rho$ and $\delta$ for fixed $\xi$ and $\Gamma_\rho/m_\rho$. Each curve corresponds to the slice at $\xi=0.05, 0.1$ or 0.2 in Fig.~\ref{fig:phase}.}
\label{fig:phase-mass}
\end{figure}
\end{center}

In Fig. \ref{fig:phase}, using Eqs.~(\ref{phase1}) and (\ref{phase2}), we can find the mass and the width of the new resonance related to the compositeness parameter $\xi$.
This figure shows that the mass of the new resonance state denoted by $m_\rho$ is converted from the unitarity limit for $\xi$.
We set that the width-to-mass ratio $\Gamma_\rho/m_\rho$ in the upper-left, upper-right and bottom-left panels is taken to be 0.1, 0.2 and 0.3, respectively.
Notice that the shaded regions in the figures are always on the unitarity circle, and the energy scale $\sqrt{S}$ is considered to be just a parameter at this stage.
As mentioned above, the non-vanishing compositeness parameter $\xi$ leads the unitarity into failure.
However, the new resonance can recover the unitarity with a suitable mass and width as shown in the figures.
There exists the maximal value of $m_\rho$ for each $\xi$ with $\delta=\pi/4$, which corresponds to $({\rm Re}[a_0],{\rm Im}[a_0])=(1/2,1/2)$ on the unitarity circle.
If $m_\rho$ exceeds the maximal value, perturbative unitarity is no longer maintained, while the broader $\Gamma_\rho$ indicates the higher $m_\rho$.
We also show the correlation between $\delta$ and $m_\rho$ in Fig.~\ref{fig:phase-mass} which corresponds to a slice with a fixed $\xi$ in the other three figures.
In the figure, dotted, solid and dashed curves represent the cases of $\xi=0.2,0.1$ and $0.05$, respectively, and the curves depicted by red, blue and orange respectively show the cases of $\Gamma_\rho/m_\rho(\equiv x)=0.1,0.2$ and $0.3$.

Next, let us discuss the possibility to observe the phase shift at the current and future hadron colliders.
As explained above, it is basically possible to extract information of the new resonance such as,
$m_\rho$ and $\Gamma_\rho$, by measuring the phase shift. 
Although the phase shift may be able to be directly extracted form the cross section measurement of $W_LW_L$ scattering, it requires very high energy collider experiments~\cite{Contino:2010mh}.
As an alternative way, we consider the full leptonic channel of $u\bar d\to WZ$ production process since the phase shift is expected to be accessible in this process.
In general, the phase shift in this process is not necessarily the same as that in $W_LW_L$ scattering process.
However, as studied in Ref.~\cite{Murayama:2014yja}, we assume that the phase shifts in both scattering processes are essentially same.
The phase shift $\delta$ induces the angular correlation between the direction of the charged lepton from the decay of the
$W$ boson and the production plane of $W$ and $Z$~\cite{Keung:2008ve,Cao:2009ah,Murayama:2014yja}, whose kinematics is explained in Appendix~\ref{sec:Kinematics}.
The angular correlation can be measured by the angular distribution of the charged lepton from the $W$ boson in $lll+\slashed{E}_T$ final states.

\begin{figure}[htbp]
\begin{center}
	\includegraphics[scale=0.7]{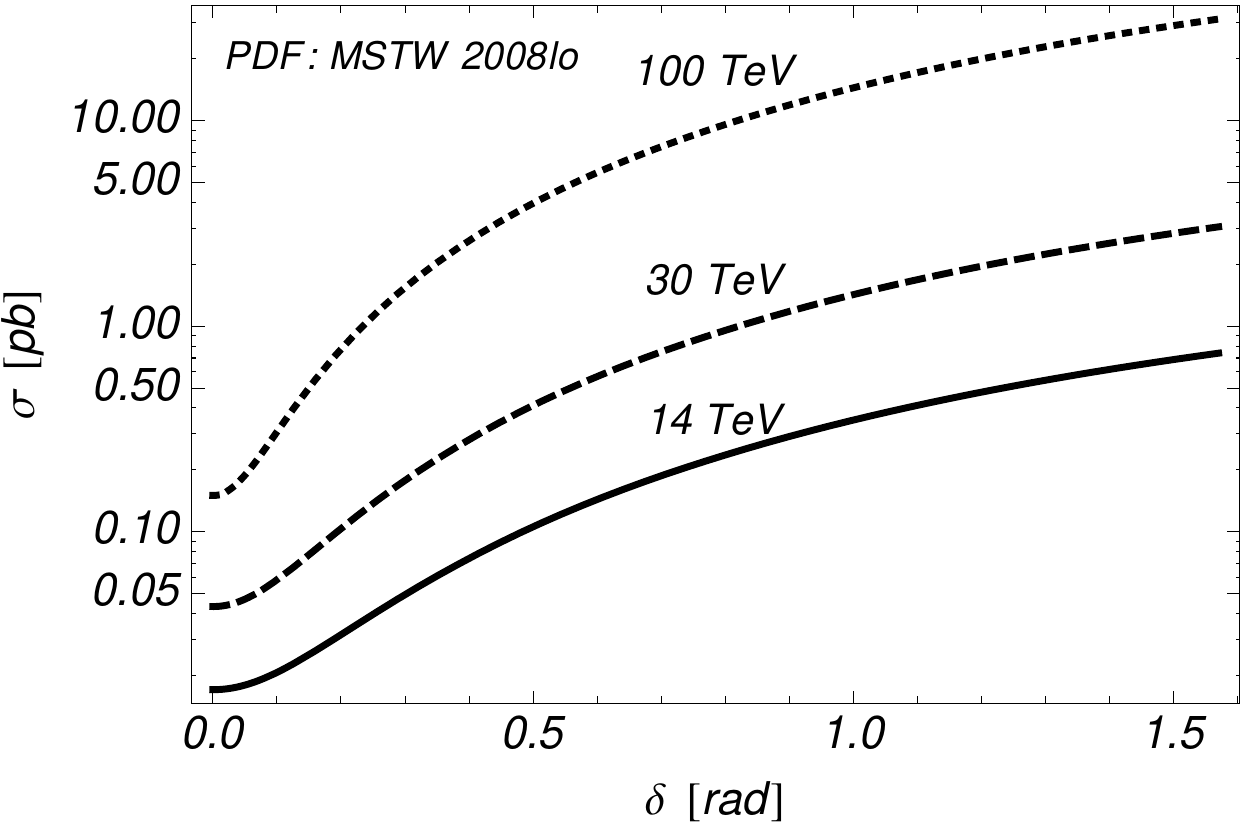}
\end{center}
\caption{Total cross section as a function of the phase $\delta$. Solid, dashed and dotted curves
show the case that the center of mass energy is 14 TeV, 30 TeV and 100 TeV, respectively. 
At the point of $\delta=0$, the cross section is given by $\sigma(14~{\rm TeV})\sim0.01~{\rm pb}$, 
$\sigma(30~{\rm TeV})\sim0.04~{\rm pb}$ and $\sigma(100~{\rm TeV})\sim0.1~{\rm pb}$.}
\label{fig:Xs}
\end{figure}

Here we define the production amplitude of $u\bar d\to WZ$ as ${\cal M_{\rm Prod}}(\Theta;\lambda_W,\lambda_Z)$ as a function of $\lambda_W$ and $\lambda_Z$, which are polarizations of $W$ and $Z$ bosons, respectively,
where $\lambda_W$ and $\lambda_Z$ are taken to be $-1$, $0$ and $+1$.
The angle $\Theta$ represents the scattering angle between incoming $u$-quark and outgoing $W^+$.
Meanwhile, amplitudes of $W^+\to l^+\nu_l$ and $Z\to l^+l^-$ are ${\cal M}_W(\theta_1,\phi_1;\lambda_W)$ 
and ${\cal M}_Z(\theta_2,\phi_2;\lambda_Z)$, respectively, depending on two polar angles $(\theta_1,\theta_2)$ 
and two azimuthal angles $(\phi_1,\phi_2)$ of the final state leptons\footnote{See Appendix A.}.
Hence the differential cross section is proportional to
\footnote{In the case of $e^+e^- \to W^+W^-$, the detailed calculation is given in Ref.~\cite{Hagiwara:1986vm}.}
\begin{equation}
d\sigma(u\bar d\to WZ\to l\nu ll)\propto
\left|
\sum_{\lambda_W,\lambda_Z}{\cal M_{\rm Prod}}(\Theta;\lambda_W,\lambda_Z){\cal M}_W(\theta_1,\phi_1;\lambda_W){\cal M}_Z(\theta_2,\phi_2;\lambda_Z)
\right| ^2.
\end{equation}
When a new resonance exists, a non-vanishing phase is induced in the scattering amplitude for $W_L$ and $Z_L$.
We then introduce the phase by ${\cal M_{\rm Prod}}(\Theta;0,0)\to {\cal M_{\rm Prod}}(\Theta;0,0) e^{i\delta}$, and we regard this $\delta$ as the phase shift which appears in $W_LW_L$ scattering.
Figure~\ref{fig:Xs} shows the $\delta$ dependence of the total cross section $\sigma(pp\to WZ\to l\nu ll)$ in $pp$ collisions, where the solid, dashed and dotted curves show 
the case that the collision energies are 14 TeV, 30 TeV and 100 TeV, respectively. 
In numerical computation we utilize the parton distribution function (PDF) of MSTW 2008lo~\cite{Martin:2009iq}.
At the point of $\delta=0$ (corresponding to the SM), the cross section is given by $\sigma(14~{\rm TeV})\sim0.01~{\rm pb}$, $\sigma(30~{\rm TeV})\sim0.04~{\rm pb}$ and $\sigma(100~{\rm TeV})\sim0.1~{\rm pb}$.

\begin{figure}[htbp]
\begin{center}
		\includegraphics[scale=0.7]{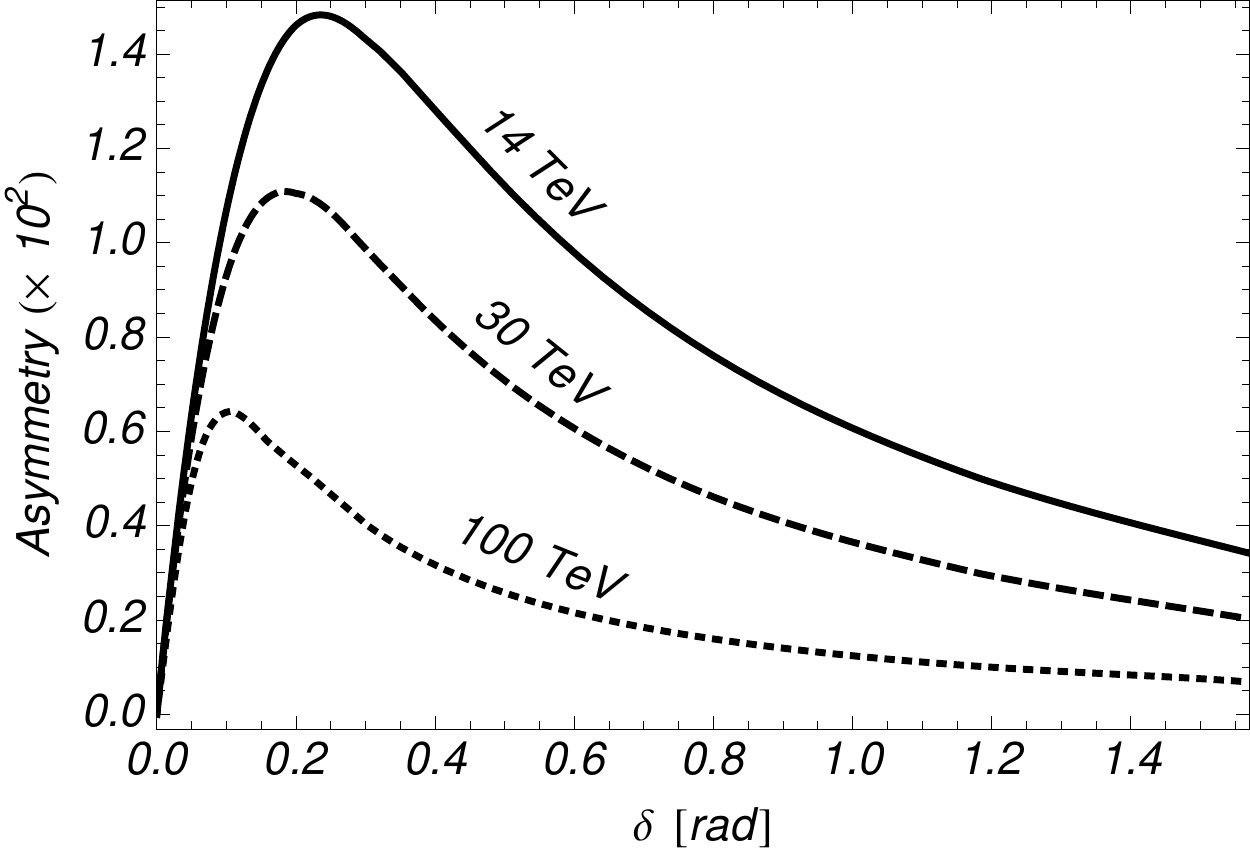}
\end{center}
\caption{The asymmetry $A_\pm$ as a function of $\delta$, 
where solid, dashed and dotted curves correspond to the case of the 
collision energy 14 TeV, 30 TeV and 100 TeV, respectively.}
\label{fig:Asymmetry}
\end{figure}

Since the decay amplitudes depend on the azimuthal angles as 
${\cal M}_W(\theta_1,\phi_1;\lambda_W)\propto e^{i\lambda_W\phi_1}$ and ${\cal M}_Z(\theta_2,\phi_2;\lambda_Z)\propto e^{-i\lambda_Z\phi_2}$, 
the squared amplitude gives interference terms between $(\lambda_W,\lambda_Z)=(0,0)$ and other transversally polarized states.
Thus, non-vanishing angular correlation such as $|{\rm amplitude}|^2\supset \sin(\lambda_W\phi_1-\lambda_Z\phi_2)\sin\delta$ is coherently induced for $\lambda_W\neq 0$ and/or $\lambda_Z\neq 0$.
Such a correlation vanishes when $\delta=0$, and only appears in the case of $\delta\neq 0$.
Therefore, the phase shift $\delta$ can be extracted by using the asymmetric behavior of the flight direction of a charged lepton with $\phi_1$ or $\phi_2$.
Hereafter we will focus on $\phi_1$ that is the azimuthal angle of the charged lepton from the $W$ decay.
Notice that there are two kinds of ambiguity to identify the events.
One is the misidentification of $u$-quark direction, which leads to $(\Theta,\phi_1,\phi_2)\to(\pi-\Theta,\pi+\phi_1,\pi+\phi_2)$.
The coefficient of $\sin\phi_1$ in the squared amplitude approximately transforms as odd.
Thus the phase space integration over $0<\cos\Theta<1$ (or $-1<\cos\Theta<0$), which picks up a contribution of one colliding direction of $u$-quark, gives a robust result.
Another issue is the missing energy of the neutrino, which makes $\phi_1$ obscure.
The asymmetry between the differential cross sections integrated over $0<\phi_1<\pi$ and $\pi<\phi_1<2\pi$ is defined by
\begin{eqnarray}
	A_{\pm} &\equiv& \frac{|\sigma_+ - \sigma_-|}{\sigma_+ + \sigma_-},
	\quad \sigma_{\pm}\equiv \sigma(\sin\phi_1 \gtrless 0),
	\label{eq:asymmetry}
\end{eqnarray}
where the cross section is given by integrating over $0<\cos\Theta<1$. 
This asymmetry is not affected by the direction of outgoing neutrino, and can be a viable quantity in the present case.
Numerically, the asymmetry is evaluated as shown in Fig.~\ref{fig:Asymmetry} as a function of $\delta$ for collision energies 
of 14 TeV, 30 TeV and 100 TeV.

\begin{figure}[htbp]
\begin{center}
		\includegraphics[scale=0.7]{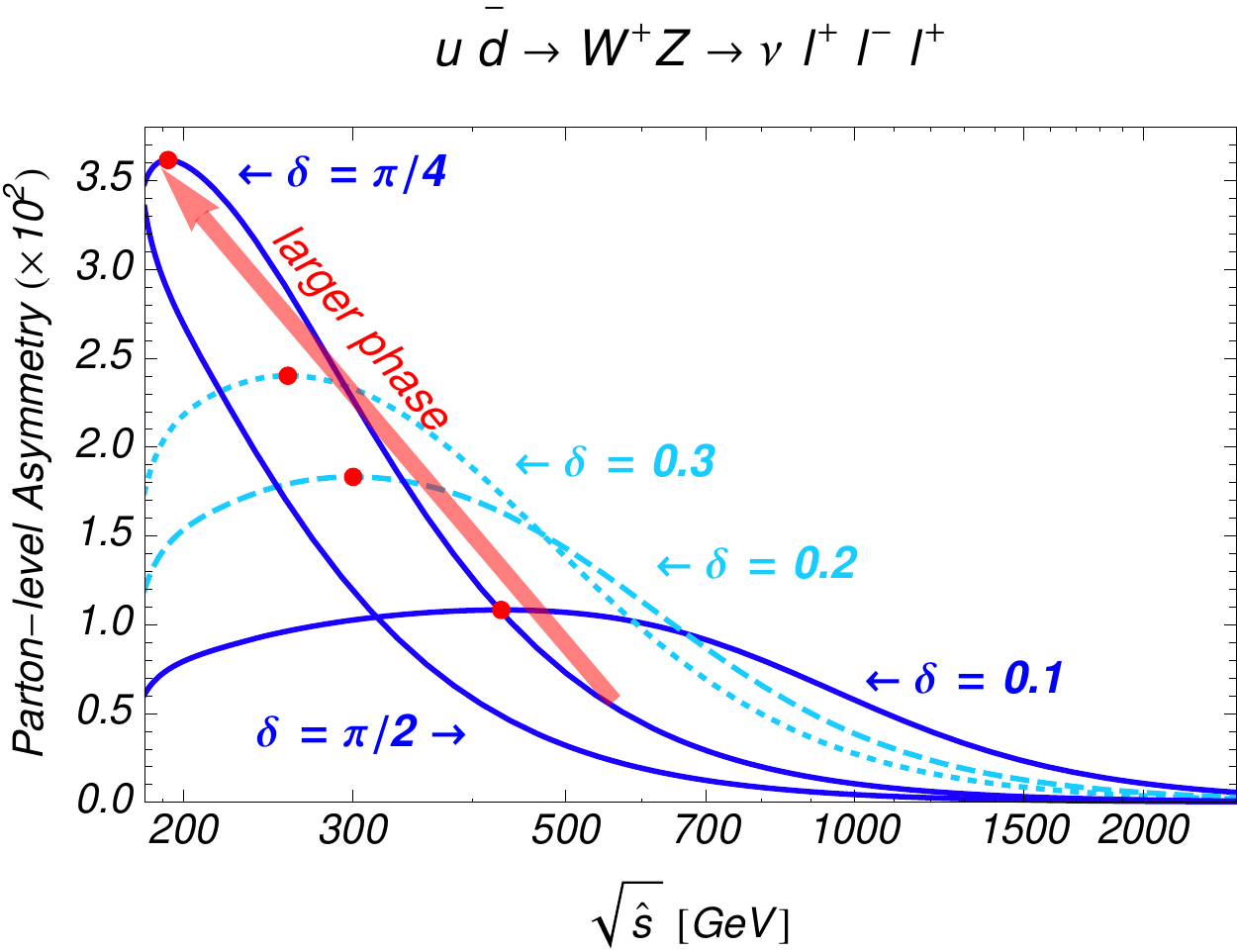}
\end{center}
\caption{Parton-level asymmetry as a function of the center of mass energy $\sqrt{\hat s}$. 
The asymmetry is shown for the case of $\delta=0.1,0.2,0.3,\pi/4$ and $\pi/2$, and 
the red dots represent the points at which the asymmetry becomes maximally large for each $\delta$.}
\label{fig:Parton-level Asymmetry}
\end{figure}

The asymmetry already appears at the parton level, which is defined by
\begin{eqnarray}
	\hat A_{\pm} &\equiv& \frac{|\hat\sigma_+ - \hat\sigma_-|}{\hat\sigma_+ + \hat\sigma_-},
	\quad \hat\sigma_{\pm}\equiv \hat\sigma(\sin\phi_1 \gtrless 0),
	\label{eq:asymmetry_parton}
\end{eqnarray}
where $\hat\sigma_\pm$ are parton-level cross sections.
The parton-level asymmetry depends on the center-of-mass energy
of partons $\sqrt{\hat s}$.
In Fig.~\ref{fig:Parton-level Asymmetry}, we show the parton-level asymmetry as a function of $\sqrt{\hat s}$ for $\delta=0.1,0.2,0.3,\pi/4$ and $\pi/2$.
The red dots in the figure represent the points at which the asymmetry becomes maximally large for each $\delta$.
A larger phase pushes down the maximal point into a lower energy.
After the convolution with the PDF, 
it can be seen that the maximal point appears around lower energy.
Eventually the asymmetry becomes large around $\delta\sim 0.2$ for the collision energy to be 14 TeV as shown in Fig.~\ref{fig:Asymmetry}.
The asymmetry becomes smaller for larger phases, which can be understood by kinematics as follows.
As shown in Fig.~\ref{fig:Parton-level Asymmetry}, when the phase is larger, the value of $\sqrt{\hat s}$ where the parton-level asymmetry becomes maximal gets smaller, and finally goes below the $WZ$ threshold.

In Fig.~\ref{fig:Asymmetry}, the asymmetry also becomes smaller when the phase is getting smaller.
This behavior is reasonable because a smaller phase shift corresponds to a higher resonance scale so that we cannot reach such an energy scale.
We find that the asymmetry is most sensitive to the case of $\delta\sim 0.2$ at LHC with the collision energy 14 TeV.
When we extend the collision energy to 30 TeV and 100 TeV, the sensitivity to a smaller phase increases, as shown by the dashed and dotted curves 
in the figure.

Before closing this section, let us discuss the possibility to observe the phase shift in this channel at future hadron collider experiments.
As shown in Fig.~\ref{fig:Xs}, the non-vanishing phase predicts larger cross section compared to the SM prediction.
The asymmetry is expected to be observed at future LHC experiments.
For example, $\sigma\simeq 0.037~{\rm pb}$ with the collision energy 14 TeV when $\delta\simeq 0.24$ (and $A_\pm\simeq 0.015$).
In this case, the statistic error is comparable to the value of the asymmetry with integrated luminosity 300 fb$^{-1}$, and thus more statistics is required.
If we accumulate 3000 fb$^{-1}$ at the high-luminosity phase of future LHC experiment, 
the statistics is almost one order of magnitude improved, and thus the phase shift can be observed.
Furthermore, when we consider higher-energy collider experiments such as collision energies 30 TeV and 100 TeV, the cross section get larger.
In the case of the collision energy 30 TeV, the cross section becomes $\sigma\simeq 0.095~{\rm fb}$ at $\delta\simeq 0.018$ where the asymmetry is maximally large.
Although the cross section with the collision energy 30 TeV becomes large as compared to that with 14 TeV, 
the maximal asymmetry decreases because $A_\pm$ is normalized by the cross section itself.
Thus the higher luminosity is also required in this case.
The sensitivity to the asymmetry is not improved even in the case of the collision energy 100 TeV.
Therefore, in order to observe the asymmetry, the increasing luminosity might be efficient rather than the increasing collision energy.

It should be noted that for the signal event $pp\to WZ\to l\nu ll$ there are large background events.
However, since we see the asymmetry, it is not necessary to suppress the SM background after extracting the events. 
The efficiency of the event selection for $pp\to WZ\to l\nu ll$ is about 70\%~\cite{ATLAS:1999vwa,Ball:2007zza}. Although this efficiency makes the statistic error larger, our naive estimation is not largely affected by the background.
An ingenious technique is also helpful to probe the phase shift in this channel.
For example, although we have studied only leptonic decay of $Z$, larger cross sections 
can be achieved by taking hadronic decay modes into account, where the asymmetry does not decrease since it is induced by the $W$ decay.

Let us finally comment on the future $e^+e^-$ collision experiments such as ILC in which more precise measurement can be achieved.
For example, the similar procedure can be applied to $e^+e^-\to W^+W^-\to l^+\nu_l \bar u d$ so that information of the phase shift can be extracted from the kinematics.
It might also be possible to apply the same manner to the Higgs-strahlung process.
These cases will be studied elsewhere.

\section{Fingerprint identification in the MCHMs}
\label{sec:fingerprint}

As discussed in the previous section, appearance of a new resonance with a relatively broad width can be the first evidence of composite Higgs scenarios.
However, there exists a variety of MCHMs depending on matter representations so that as the second step we have to narrow down the MCHMs to a class of specific models by experiments.
One of the promising strategies for this purpose is to fingerprint MCHMs by precisely measuring a set of the Higgs boson couplings.
The precision measurements at the high-luminosity LHC as well as at future $e^+e^-$ colliders will be able to provide 
a strong clue to understand the detail of the MCHMs.
In this section, we demonstrate how to distinguish variations of the MCHMs by patterns of the deviations from the SM predictions.
In order to investigate such deviations, 
we utilize scale factors defined by $\kappa_a\equiv g_a/g_a^{\rm SM}$, where 
$g_a$ denote the Higgs boson couplings with the weak gauge bosons $V(V=W,Z)$, matter fermions 
and the Higgs boson itself such as $a=hVV$, $hhh$, $htt$ and $hbb$.
For some of them, we use simple forms as $\kappa_V\equiv \kappa_{hVV}$, 
$\kappa_t\equiv \kappa_{htt}$, and $\kappa_{b}\equiv \kappa_{hbb}$.
We also discuss contact interactions
such as $hhVV$, $hhhh$, $hht\bar t$ and $hhb\bar b$, where we define their couplings as $g_{hhVV}^{}$, $g_{hhhh}^{}$, $g_{hhtt}^{}$ and $g_{hhbb}^{}$.
For $hhVV$ and $hhhh$, we use the parameters $c_{hhVV}^{}\equiv g_{hhVV}^{}/g_{hhVV}^{\rm SM}$ and $c_{hhhh}^{}\equiv g_{hhhh}^{}/g_{hhhh}^{\rm SM}$.
Each MCHM basically predicts a specific pattern of deviations in these couplings, so that we can distinguish models by detecting 
such a pattern by experiments.

As already shown in Sec.~\ref{sec:review}, 
universal predictions for the scale factors of the Higgs couplings to the gauge bosons 
are obtained as $\kappa_V^{}=\sqrt{1-\xi}$ and $c^{}_{hhVV}=1-2\xi$ in the MCHMs.
It means that the compositeness parameter $\xi$ is determined by the measurement of $\kappa^{}_V$.
We can also test the consistency with the MCHMs by measuring the correlations among $\kappa^{}_V$ and $c^{}_{hhVV}$ independent of the detail in matter sector of the MCHMs.
For example, in the minimal supersymmetric SM, $\kappa^{}_V$ is reduced by the mixing angle, but $c^{}_{hhVV}$ is always unity regardless of the mixing angle.
However, it could be challenging to precisely measure $c^{}_{hhVV}$ even at future collider experiments 
and should be a task for future colliders~\cite{Killick:2013mya}.

On the other hand, the main contribution to the one-loop effective Higgs potential is driven by the Yukawa coupling of matter fermions.
Therefore, self-couplings of the Higgs boson as well as the Higgs boson couplings to the matter 
fermions reflect the matter sector of the MCHMs $\mathcal{L}_{\text{eff}}^{\text{matter}}$.
The effective Lagrangian for the matter sector is determined by how the SM matter fermions are 
embedded into the $SO(5)$ representations.
In the following, we define various MCHMs according to the matter representations in order, and discuss deviations in Higgs boson couplings.

First, we introduce the simplest model, so-called $\text{MCHM}_4$. 
In this model, all the matter fermions are embedded into 
four-dimensional representations $\Psi_r^{(4)}(r=q,u,d)$ of $SO(5)$ as 
\begin{equation}
	\Psi_q^{(4)}=\begin{pmatrix}
		q_L\\
		Q_L
	\end{pmatrix}\;,\quad
	\Psi_u^{(4)}=\begin{pmatrix}
		q_R^u\\
		\begin{pmatrix}
			u_R\\
			d_R^{\prime}
		\end{pmatrix}
	\end{pmatrix}\;,\quad
	\Psi_d^{(4)}=\begin{pmatrix}
		q_R^d\\
		\begin{pmatrix}
			u_R^{\prime}\\
			d_R
		\end{pmatrix}
	\end{pmatrix}\;.
\end{equation}
Here, $q_L=(u_L,d_L)^T$, $u_R$, and $d_R$ are SM quark $SU(2)_L$ doublet, right-handed up-type quark and 
right-handed down-type quark, respectively, and 
the other fields as $Q_L$, $q_R^u$, $q_R^d$, $u_R^{\prime}$, and $d_R^{\prime}$ are non-dynamical fields so that 
their contributions are negligible. 
The relevant matter part of the effective Lagrangian is given by 
\begin{equation}
	\mathcal{L}_{\text{eff}}^{\text{matter}}=
	\sum_{r=q,u,d}\overline{\Psi}_r^{(4)}\slashed{p}\left[\Pi_0^r(p)+\Pi_1^r(p)\Gamma^i\Sigma_i\right]\Psi_r^{(4)}
	+
	\sum_{r=u,d}\overline{\Psi}_q^{(4)}\left[M_0^r(p)+M_1^r(p)\Gamma^i\Sigma_i\right]\Psi_r^{(4)}\;,
	\label{eq:matterLag}
\end{equation}
where $\Gamma^i(i=1,\cdots,5)$ are gamma matrices in five-dimensional representation of $SO(5)$, and $M$'s are 
the form factors.
The loop contributions of the matter fermion to the Higgs potential is dominated by the 
top-quark loop, and it is evaluated in the $\text{MCHM}_4$ as 
\begin{equation}
	V_{\text{eff}}^{\text{fermion}}\simeq -2N_C\int\frac{d^4p}{(2\pi)^4}
	\left[\ln \slashed{p}\Pi_{b_L}+\ln (p^2\Pi_{t_R}\Pi_{t_L}-\Pi_{t_Lt_R}^2)\right]\;,
	\label{eq:vfermion}
\end{equation}
where 
\begin{align}
	\Pi_{t_L}=\Pi_{b_L}=\Pi_0^q+\Pi_1^q\sin(h/f)\;,\;\;
	\Pi_{t_R}=\Pi_0^u-\Pi_1^u\cos(h/f)\;,\;\;
	\Pi_{t_Lt_R}=M_1^u\sin(h/f)\;,
\end{align}
and $N_C=3$ is the colour number of QCD.
Notice that this contribution $V_{\text{eff}}^{\text{fermion}}$ depends on the representation of 
the quark fields.
From Eqs.~(\ref{eq:vgauge}) and (\ref{eq:vfermion}),
the effective potential given in Eq.~(\ref{eq:HiggsPotential0}) can be rewritten as
\begin{equation}
	V_{\text{eff}}\simeq \alpha \cos(h/f)-\beta\sin^2(h/f)\;,
\end{equation}
where 
\begin{align}
	\alpha =& 2N_C\int\frac{d^4p}{(2\pi)^4}\left(\frac{\Pi_1^u}{\Pi_0^u}-2\frac{\Pi_1^q}{\Pi_1^q}\right)\;,\nonumber\\
	\beta =& \int\frac{d^4p}{(2\pi)^4}\left(2N_C\frac{|M_1^u|^2}{(-p^2)(\Pi_0^q+\Pi_1^q)(\Pi_0^u-\Pi_q^u)}-\frac{9}{8}\frac{\Pi_1^{}}{\Pi_0^{}}\right)\;.
\end{align}
By the contribution of $V_{\text{eff}}^{\text{fermion}}$, the $SU(2)_L\times U(1)_Y$ is broken at the minimum of the effective potential $V_{\text{eff}}$.
Actually, the vacuum conditions given by, 
\begin{align}
	&\left\langle\frac{\partial V_h}{\partial h}\right\rangle =\frac{\sin(\langle h\rangle /f)}{f}\left(\alpha + \beta\cos(\langle h\rangle /f)\right)=0\;,\nonumber\\
	&\left\langle\frac{\partial^2 V_h}{\partial h^2}\right\rangle =\frac{2\beta}{f^2}\left(1-\frac{\alpha^2}{4\beta^2}\right)=m_h^2>0\;,
\end{align}
are satisfied with $\sin(\langle h\rangle /f)=v/f \neq 0$.
The coupling constant for the triple Higgs boson coupling is predicted as 
\begin{equation}
	\lambda_{hhh}\equiv\left\langle\frac{\partial^3 V_h}{\partial h^3}\right\rangle =\frac{3m_h^2}{v}\sqrt{1-\xi}\;.
\end{equation}

Eq.~(\ref{eq:matterLag}) also leads to 
the mass terms of the third generation quarks and these interaction terms with the Higgs boson as 
\begin{align}
	\mathcal{L}_{\text{eff}}=&M_1^t\sin(h/f)\bar{t}t+M_1^b\sin(h/f)\bar{b}b\nonumber\\
	=&M_1^t\sqrt{\xi}\left(1+\sqrt{1-\xi}\frac{\hat{h}}{v}-\frac{1}{2}\xi\frac{\hat{h}^2}{v^2}+\cdots\right)\bar{t}t
	+M_1^b\sqrt{\xi}\left(1+\sqrt{1-\xi}\frac{\hat{h}}{v}-\frac{1}{2}\xi\frac{\hat{h}^2}{v^2}+\cdots\right)\bar{b}b\nonumber\\
	=&m_t\bar{t}t +\frac{m_t}{v}\sqrt{1-\xi}\hat{h}\bar{t}t-\frac{m_t}{2v^2}\xi \hat{h}^2\bar{t}t
	+m_b\bar{b}b +\frac{m_b}{v}\sqrt{1-\xi}\hat{h}\bar{b}b-\frac{m_b}{2v^2}\xi \hat{h}^2\bar{b}b+\cdots\;,
\end{align}
where $m_t$ and $m_b$ are the masses of the top quark and the bottom quark, respectively.
It provides us $\kappa_t=\kappa_b=\sqrt{1-\xi}$.
For the contact interactions of two Higgs bosons and two fermions, their coupling constants are given by $g_{hhtt}=-m_t\xi/(2v^2)$ and $g_{hhbb}=-m_b\xi/(2v^2)$ in the $\text{MCHM}_4$ model.
We parametrize these couplings as $c_{hhtt}\equiv g_{hhtt}/(m_t/(2v^2))$ and $c_{hhbb}\equiv g_{hhbb}/(m_b/(2v^2))$ in the following discussions.

Next, we consider variations of MCHMs.
There are other representations of $SO(5)$ into which we can embedd the SM quark fields, such as one-, five-, ten- and fourteen-dimensional representations and so on.
In general, $q_L=(u_L,d_L)$, $u_R$, and $d_R$ can be embedded into individual representations.
We already discussed one of the simplest model, MCHM$_4$.
Another simple model is the $\text{MCHM}_5$ in which all the quark fields, $u_L$, $d_L$, $u_R$ and $d_R$, are embedded into the five-dimensional representations.
The detail of the model is given in the Appendix~\ref{sec:models}.
The factors $\kappa_{V}^{}$ and $c_{hhVV}^{}$, in the $\text{MCHM}_5$ 
are the same as those in the $\text{MCHM}_4$. 
On the other hand, the $\text{MCHM}_5$ predicts different deviation patterns  
from the $\text{MCHM}_4$ predictions for the factors
$\kappa_{hhh}^{}$, $c_{hhhh}^{}$, $\kappa_t^{}$, $\kappa_b^{}$ $c_{hhtt}^{}$ and $c_{hhbb}^{}$.
The $\text{MCHM}_4$ and $\text{MCHM}_5$ predict 
$(\kappa_{hhh}^{}, c_{hhhh}^{}, \kappa_{t(b)}^{}, c_{hhtt(hhbb)}^{})\simeq(1-\frac{1}{2}\xi, 1-\frac{7}{3}\xi, 1-\frac{1}{2}\xi, -\xi)$ and
$(1-\frac{3}{2}\xi, 1-\frac{25}{3}\xi, 1-\frac{3}{2}\xi, -4\xi)$, respectively.
If these deviations can be measured prescisely enough, we can 
distinguish the $\text{MCHM}_4$ and the $\text{MCHM}_5$.

Similarly, we can classify the MCHMs by the precision measurements
of the deviation patterns in the Higgs boson couplings.
In order to demonstrate the classification of the MCHMs, 
we consider several models with different representations
of the quark fields.
The MCHMs discussed here and the predicted deviation patterns are listed in the Table~\ref{tab1}.
The effective Lagrangian for the matter sector  
and the Higgs potential in each models are shown in the Appendix~\ref{sec:models}.
In the model named $\text{MCHM}_{i\text{-}j\text{-}k}$, 
the quark fields $q_L=(u_L,d_L)$, $u_R$, $d_R$ are embedded into $i$-, $j$- and $k$-dimensional representations, respectively.
In the case of $i=j=k$, we simply write $\text{MCHM}_i$ instead 
of $\text{MCHM}_{i\text{-}i\text{-}i}$.
Patterns of scale factors in various models are partly studied in Ref.~\cite{Carena:2014ria}.
In this paper, we make more complete list of the models\footnote{
	We cannot make a realistic model for some combinations of the matter representations.
	For example, in the model $\text{MCHM}_\text{5-1-10}$, the electroweak symmetry breaking cannot occur 
	as shown in the Appendix~\ref{sec:models}.
	Therefore we don't consider such a model in the analysis of the scale factors.
}
and we add the predictions on the deviations for additional intereractions such as $hhVV$, $hhhh$, $hhtt$, and $hhbb$.
In the table, we use the functions defined in 
Ref.~\cite{Carena:2014ria} as
\begin{eqnarray}
	F_3 &=& \frac{1}{\sqrt{1-\xi}}\frac{3(1-2\xi)M_1^t+2(4-23\xi+20\xi^2)M_2^t}{3M_1^t+2(4-5\xi)M_2^t},\nonumber\\ 
    F_4 &=& \sqrt{1-\xi}\frac{M_1^t+2(1-3\xi)M_2^t}{M_1^t+2(1-\xi)M_2^t},\quad
    F_5 = \sqrt{1-\xi}\frac{M_1^t-(4-15\xi)M_2^t}{M_1^t-(4-5\xi)M_2^t},
\end{eqnarray}
where $M_1^t$ and $M_2^t$ are form factors in effective theories 
shown in the Appendix~\ref{sec:models}, and they cannot be determined within the 
framework of the low energy theories. 
We here additionally introduce 
\begin{eqnarray}
    F_6 &=& -4\xi\frac{3M_1^t+(23-40\xi)M_2^t}{3M_1^t+2(4-5\xi)M_2^t},\quad
    F_7 = -\xi\frac{M_1^t+2(7-9\xi)M_2^t}{M_1^t+2(1-\xi)M_2^t},\nonumber\\
    F_8 &=& -\xi\frac{M_1^t-(34-45\xi)M_2^t}{M_1^t-(4-5\xi)M_2^t},\nonumber\\
    H_1 &=& 1-\frac{3\xi}{2}-\frac{5\xi^2}{8}+\frac{\xi^3}{3m_h^2}\left[-\frac{21m_h^2}{16}+\frac{48\gamma}{v^2}\right],\nonumber\\
    H_2 &=& 1-\frac{25\xi}{2}+\xi^2+\frac{\xi^3}{3m_h^2}\left[3m_h^2+\frac{288\gamma}{v^2}\right],
\end{eqnarray}
where $\gamma$ is one of the form factors defined in the Appendix~\ref{sec:models}.
In the models such that two different form factors $M_1^t$ and $M_2^t$ contribute to 
the scale factor, we examine two typical cases of $M_1^t\to 0$ or $M_2^t\to 0$ for simplicity.
For the form factors $H_1$ and $H_2$, we take the terms up to $O(\xi^2)$, and 
thus the contribution through the $\gamma$ term in the potential Eq.~(\ref{eq:Vh14}) can be neglected 
because it is proportional to $\xi^3$.

\begin{center}
\begin{figure}
	\begin{tabular}{cc}
		\includegraphics[width=80mm]{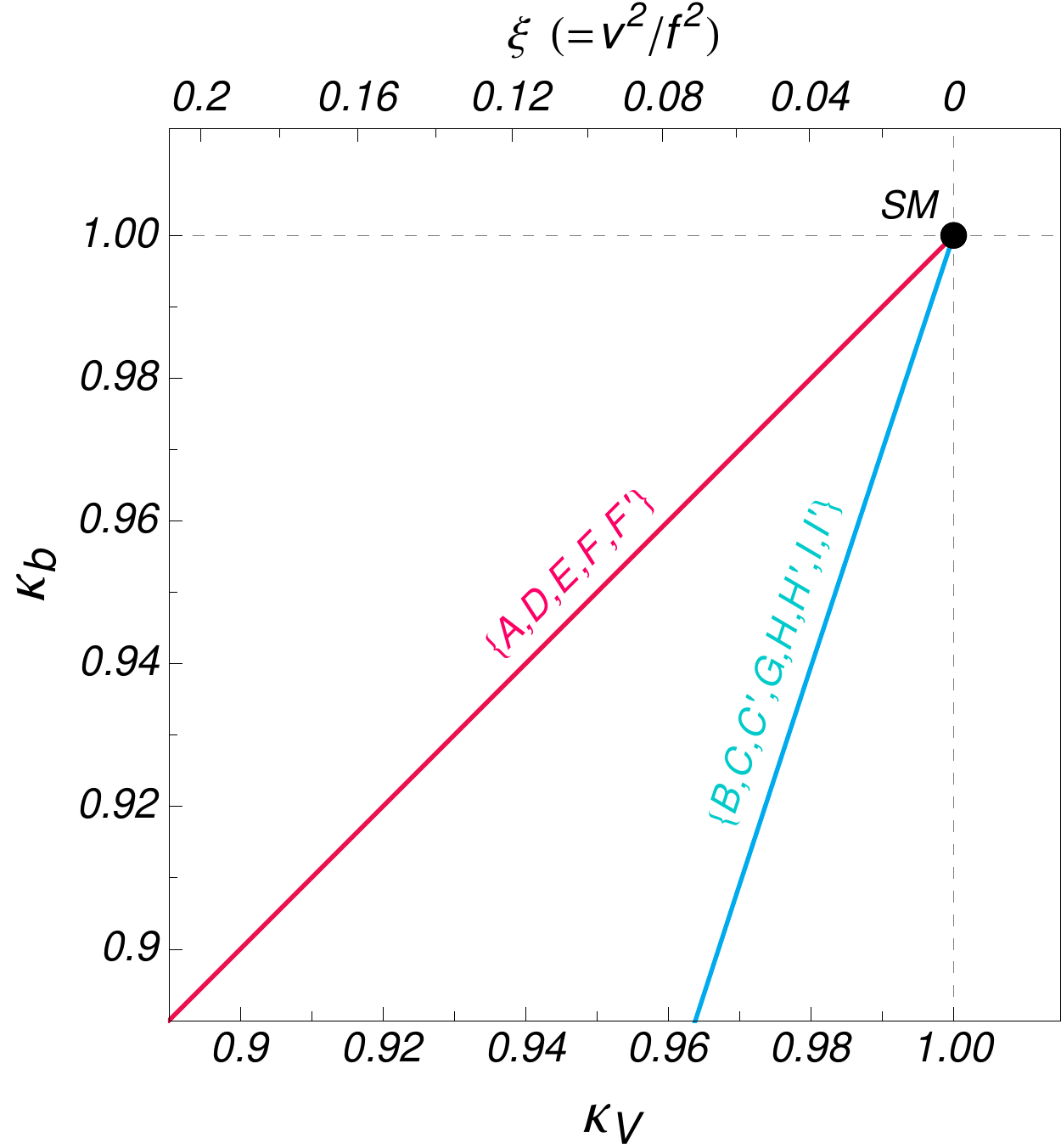}&
		\includegraphics[width=80mm]{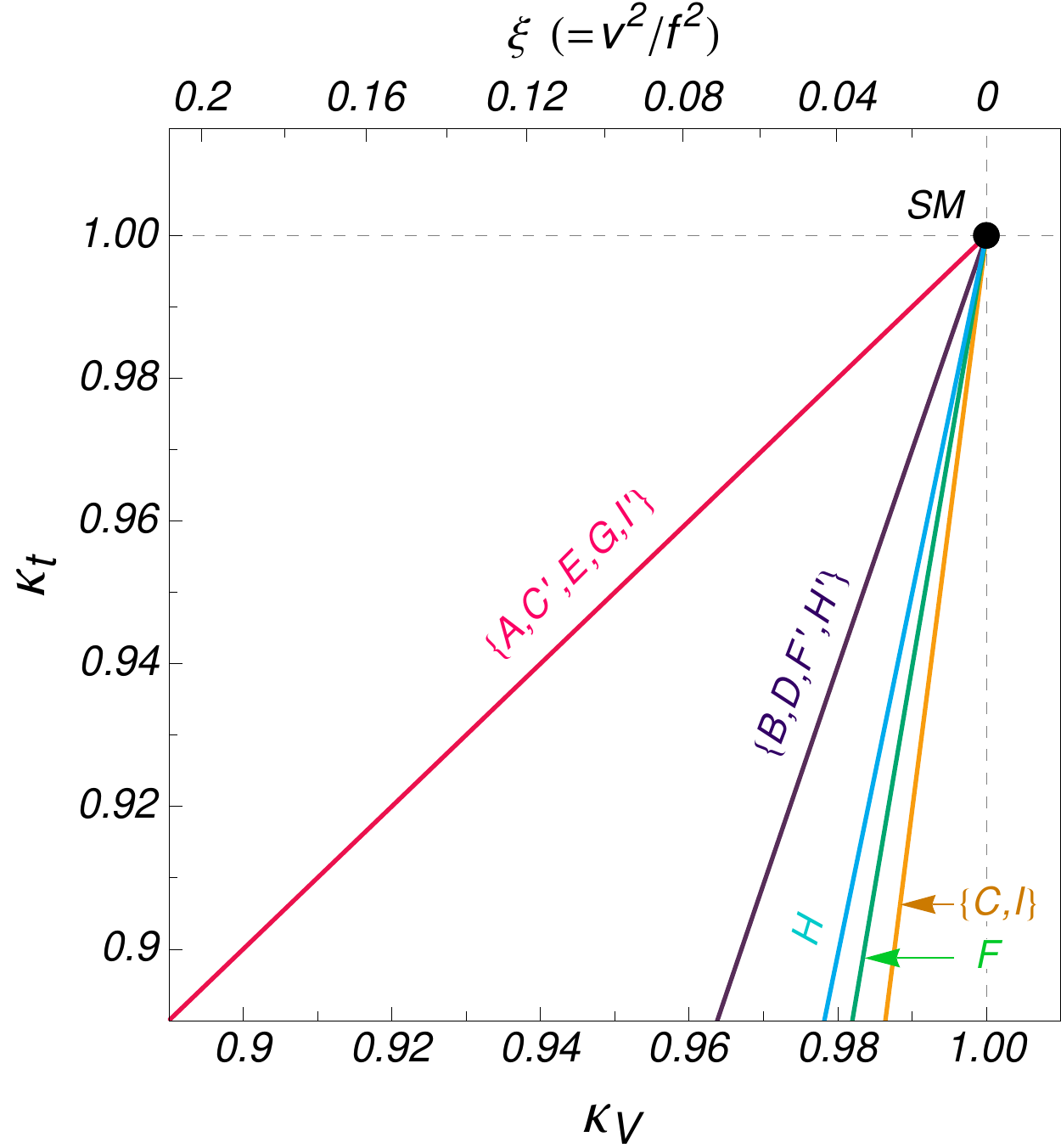}\\
	\includegraphics[width=85mm]{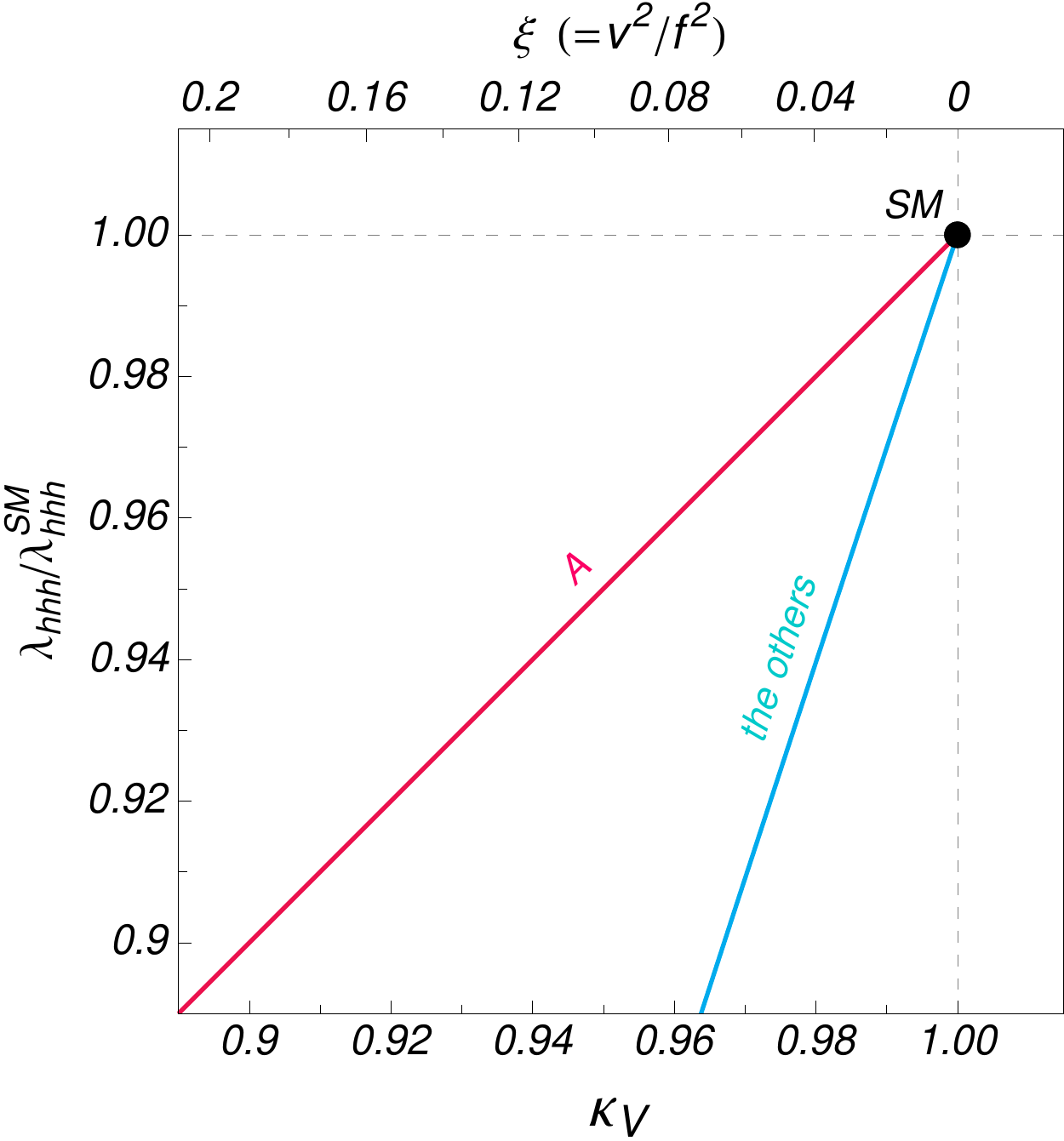}&
\end{tabular}
	\caption{Correlations between scale factors. $\kappa_V$ is universal for all MCHMs, and thus $\xi$-dependence is the same.}
	\label{couplings}
\end{figure}
\end{center}

In Fig.~\ref{couplings}, several scale factors are shown as a function of $\kappa_{V}$ which is uniquely determined by $\xi$.
Upper-left panel of Fig.~\ref{couplings} is also shown in Ref.~\cite{Giardino:2013bma} for MCHM$_4$ and MCHM$_5$\footnote{They also show the case of MCHM$_{5\text{-}1\text{-}x}$ ($x$ is arbitrary) with additional fermionic resonances~\cite{Pomarol:2012qf}. Without such resonances, the model cannot maintain electroweak symmetry breaking as is the case with MCHM$_{5\text{-}1\text{-}10}$ mentioned in Appendix B.}, and our result is consistent with their one.
As seen in the set of figures, we can basically discriminate some models from the others by the correlations among
scale factors. 
For instance, 
the models $\{$A, D, E, F, F'$\}$, which are defined in Tab.~\ref{tab1}, can be separated from the other models by $\kappa_b$. 
These four models are then classified into three sets as 
$\{$A, E$\}$,
$\{$D, F'$\}$
and 
F
by measuring $\kappa_t$.
The degeneracy between A and E can be solved by the measurement of 
$\lambda_{hhh}/\lambda_{hhh}^{\text{SM}}$.

\begin{center}
\begin{table}[htbp]
\caption{Scale factors for MCHMs with various matter representations. The labels are used in Fig.~\ref{couplings}, where C, H and I are the case of $M^t_1\to 0$, and C', H' and I' are the case of $M^t_2\to 0$.}
\label{tab1}
\begin{tabular}{c|c|c|c|c|c||c|c|c|c|c|c|c|}
 \hline
Label & Model              & $\kappa_{V}$  & $c_{hhVV}$ & $\kappa_{hhh}$ & $c_{hhhh}$ & $\kappa_{t}$ & $\kappa_{b}$ & $c_{hhtt}$ & $c_{hhbb}$\\ \hline \hline
A &
${\rm MCHM}_{4}$                                                                                                                                                                                                                         
                           & $\sqrt{1-\xi}$ & $1- 2\xi$ & $\sqrt{1-\xi}$ & $1- \frac{7}{3} \xi$& $\sqrt{1-\xi}$ & $\sqrt{1-\xi}$ & $-\xi$ & $-\xi$ \\ \hline 
B &                           
${\rm MCHM}_{5}$                                                                                                                                             
                           & $\sqrt{1-\xi}$ & $1- 2\xi$ & $\frac{1-2\xi}{\sqrt{1-\xi}}$ & $\frac{1-28 \xi/3 + 28\xi^2/3 }{1-\xi}$ & $\frac{1-2\xi}{\sqrt{1-\xi}}$ & $\frac{1-2\xi}{\sqrt{1-\xi}}$ & $-4\xi$ & $-4\xi$ \\ \hline 
B &
${\rm MCHM}_{10}$                                                                                                                                            
                           & $\sqrt{1-\xi}$ & $1- 2\xi$ & $\frac{1-2\xi}{\sqrt{1-\xi}}$ & $\frac{1-28 \xi/3 + 28\xi^2/3 }{1-\xi}$ & $\frac{1-2\xi}{\sqrt{1-\xi}}$ & $\frac{1-2\xi}{\sqrt{1-\xi}}$ & $-4\xi$ & $-4\xi$ \\ \hline 
C, C' &
${\rm MCHM}_{14}$                                                                                                                                           
                           & $\sqrt{1-\xi}$ & $1- 2\xi$ & $H_1$ & $H_2$ &$F_3$ & $\frac{1-2\xi}{\sqrt{1-\xi}}$ & $F_6$ & $-4\xi$ \\ \hline \hline
D &
${\rm MCHM}_{5 \mathchar`- 5 \mathchar`- 10}$                                                                                                                                  
                           & $\sqrt{1-\xi}$ & $1- 2\xi$ & $\frac{1-2\xi}{\sqrt{1-\xi}}$ & $\frac{1-28 \xi/3 + 28\xi^2/3 }{1-\xi}$ & $\frac{1-2\xi}{\sqrt{1-\xi}}$ & $\sqrt{1-\xi}$ & $-4\xi$ & $-\xi$ \\ \hline 
E &
${\rm MCHM}_{5 \mathchar`- 10 \mathchar`- 10}$                                                                                                                                  
                           & $\sqrt{1-\xi}$ & $1- 2\xi$ & $\frac{1-2\xi}{\sqrt{1-\xi}}$ & $\frac{1-28 \xi/3 + 28\xi^2/3 }{1-\xi}$ & $\sqrt{1-\xi}$ & $\sqrt{1-\xi}$ & $-\xi$ & $-\xi$ \\ \hline 
F, F' &
${\rm MCHM}_{5 \mathchar`- 14 \mathchar`- 10}$                                                                                                                                  
                           & $\sqrt{1-\xi}$ & $1- 2\xi$ & $H_1$ & $H_2$ & $F_5$ & $\sqrt{1-\xi}$ & $F_8$ & $-\xi$ \\ \hline \hline 
G &
${\rm MCHM}_{10 \mathchar`- 5 \mathchar`- 10}$                                                                                                                                  
                           & $\sqrt{1-\xi}$ & $1- 2\xi$ & $\frac{1-2\xi}{\sqrt{1-\xi}}$ & $\frac{1-28 \xi/3 + 28\xi^2/3 }{1-\xi}$ & $\sqrt{1-\xi}$ & $\frac{1-2\xi}{\sqrt{1-\xi}}$ & $-\xi$ & $-4\xi$ \\ \hline 
B &
${\rm MCHM}_{10 \mathchar`- 14 \mathchar`- 10}$                                                                                                                                  
                           & $\sqrt{1-\xi}$ & $1- 2\xi$ & $H_1$ & $H_2$ & $\frac{1-2\xi}{\sqrt{1-\xi}}$ & $\frac{1-2\xi}{\sqrt{1-\xi}}$ & $-4\xi$ & $-4\xi$ \\ \hline \hline 
B &
${\rm MCHM}_{14 \mathchar`- 1 \mathchar`- 10}$                                                                                                                                  
                           & $\sqrt{1-\xi}$ & $1- 2\xi$ & $\frac{1-2\xi}{\sqrt{1-\xi}}$ & $\frac{1-28 \xi/3 + 28\xi^2/3 }{1-\xi}$ & $\frac{1-2\xi}{\sqrt{1-\xi}}$ & $\frac{1-2\xi}{\sqrt{1-\xi}}$ & $-4\xi$ & $-4\xi$ \\ \hline 
H, H' &
${\rm MCHM}_{14 \mathchar`- 5 \mathchar`- 10}$                                                                                                                                  
                           & $\sqrt{1-\xi}$ & $1- 2\xi$ & $H_1$ & $H_2$ & $F_4$ & $\frac{1-2\xi}{\sqrt{1-\xi}}$ & $F_7$ & $-4\xi$ \\ \hline 
B &
${\rm MCHM}_{14 \mathchar`- 10 \mathchar`- 10}$                                                                                                                                  
                           & $\sqrt{1-\xi}$ & $1- 2\xi$ & $H_1$ & $H_2$ & $\frac{1-2\xi}{\sqrt{1-\xi}}$ & $\frac{1-2\xi}{\sqrt{1-\xi}}$ & $-4\xi$ & $-4\xi$ \\ \hline 
I, I' &
${\rm MCHM}_{14 \mathchar`- 14 \mathchar`- 10}$                                                                                                                                  
                           & $\sqrt{1-\xi}$ & $1- 2\xi$ & $H_1$ & $H_2$ & $F_3$ & $\frac{1-2\xi}{\sqrt{1-\xi}}$ & $F_6$ & $-4\xi$ \\ \hline \hline
\end{tabular}
\end{table}
\end{center}

\section{Discussion}
\label{sec:discussion}

We here discuss the strategy to distinguish MCHM from the other models which are not only the other MCHMs but also various alternative Higgs models such as the MSSM and other extended Higgs sectors.

First, suppose a new resonance exists at $m_\rho\sim 3$ TeV, the specific value of the phase shift determines the compositeness parameter $\xi$ and/or the width.
When we observe $\delta\sim 0.2$ by measuring $A_\pm$ at a future collider experiment, the compositeness parameter would be determined as $\xi\sim 0.2$ if the width is narrow enough.
The scaling factors in MCHM$_4$ and MCHM$_5$ are then predicted as $(\kappa_{hhh}-1, c_{hhVV}-1)\sim(-0.11,-0.4)$ and $(-0.33,-0.4)$, respectively.
These deviations enhance the cross section of double Higgs boson production $pp\to hhjj$ by the factor of 4.5 and 6, respectively~\cite{Haba:2013xla}.
Such enhancements can be measured and thus can be discriminated at the high luminosity LHC.
At the ILC, due to the deviation of $\kappa_{hhh}$ and $c_{hhVV}$, the cross section of $e^+e^-\to\nu\bar\nu hh$ for MCHM$_4$ (MCHM$_5$) is about 3.9 (5) times larger than the SM prediction with the collision energy 1 TeV~\cite{Haba:2013xla}.
We expect that, by this kind of analyses, each MCHM can be mostly discriminated from the others.

In models with extended Higgs sectors, the Higgs boson coupling constants can also be modified the SM values due to the effect of field mixings.
The MSSM is a good example where the Higgs sector is composed of two doublet fields.
The $hVV$ coupling constant is multiplied by the mixing factor $\sin(\beta-\alpha)$, while the up-type (down-type) Yukawa coupling is modified by $\cos\alpha/\sin\beta ~(-\sin\alpha/\cos\beta)$, where $\alpha$ is the mixing angle between CP-odd Higgs bosons and $\tan\beta$ is the ratio of vacuum expectation values of the two Higgs doublets.
Hence, in the SM-like region where $\sin(\beta-\alpha)$ $(=\kappa_V)$ is slightly smaller than unity, $\kappa_b>1$ and $\kappa_t<1$ are predicted in the MSSM.
Since both are less than unity in the MCHMs, the MSSM and MCHMs can easily be separated.

In general two Higgs doublets models (2HDMs) with the softly broken discrete symmetry for avoiding flavor changing neutral current, there are four types of Yukawa interactions; i.e., Type-I, Type-II, Type-X and 
Type-Y~\cite{Barger:1989fj,Grossman:1994jb,Aoki:2009ha,Kanemura:2014bqa}.
The Higgs sector of the MSSM is the Type-II 2HDM.
For the other types of 2HDMs, similar argument can be used for the separation from the MCHMs except for the Type-I and Type-X with $\cos(\beta-\alpha)$ to be negative.
In these models, directions of the deviations in $\kappa_b$ and $\kappa_t$ are both negative.
In order to discriminate MCHMs from these models, we need to utilize the other coupling constants than the Yukawa couplings, such as $c_{hhVV}$.
In the MCHMs, we have $c_{hhVV}=1-2\xi$ which corresponds to $c_{hhVV}=0.6$ for $\xi=0.22$, while $c_{hhVV}$ is unchanged in the 2HDMs.
Therefore, by measuring the cross section of double Higgs boson production at the high luminosity LHC and the ILC~\cite{Dolan:2012ac,Asakawa:2010xj,Hespel:2014sla,Killick:2013mya,Pierce:2006dh,Kanemura:2008ub,Barger:2003rs,Espinosa:2010vn,Grober:2010yv,Contino:2012xk,Gillioz:2012se,Haba:2013xla}, we can separate MCHMs from all the type of the 2HDMs as long as the deviation in $\kappa_{V}$ is not too small.

Finally, we mention the possibility of discriminating MCHMs from the Higgs sector with an additional singlet, where $\kappa_V, \kappa_t$ and $\kappa_b$ are reduced with the same factor $\cos\theta$ where $\theta$ is the mixing angle between the SM Higgs field and the singlet field~\cite{Schabinger:2005ei,O'Connell:2006wi,Kusenko:2006rh,BahatTreidel:2006kx,Barger:2007im}.
There is no difference in direction of the deviations in MCHMs.
However, in the model with the singlet, the reduction patterns of $c_{hhVV}$ and $\kappa_{hhh}$ are different from those in MCHMs.
Therefore, we may be able to discriminate MCHMs from the model with the singlet if the double Higgs boson production can be measured accurately enough~\cite{Bowen:2007ia,Chen:2014ask}.

In conclusion, the double Higgs boson production might be another pathway 
to answer the question; is the Higgs boson elementary or composite?

Up to now, we did not take into account the use of the appearance of higher dimensional operators such as $hhtt$ and $hhbb$.
They can also affect the double Higgs boson production via gluon fusion.
At the LHC and future hadron colliders, the dominant contribution is the gluon fusion process, which is induced by top-quark loop at leading order\footnote{The vector boson fusion and the Higgs-strahlung processes also exist at hadron colliders. However, the cross section of $pp\to jjhh$ is about 10 times smaller than the gluon fusion process in the SM.}.
Therefore, the cross section is sensitive to not only $\kappa_{hhh}$ but also $c_{hhtt}$.
In particular, the latter is important in $pp\to t\bar t hh$ as well, which will be a crucial target at future hadron collider experiments~\cite{Englert:2014uqa}.
We shall examine this point as our next task.

\section{Summary}
\label{sec:summary}
In this paper, we have discussed two complementary approaches to investigate composite nature of the Higgs boson in MCHMs.
These two approaches can lead an answer to the big question; is the Higgs boson composite or elementary?

The first step which we have discussed is to extract information of the new resonance scale from the possible phase shift in longitudinally polarized vector boson pair production at hadron collider experiments.
According to the analogy with pion-pion scattering processes, a sizable phase shift can be predicted to restore perturbative unitarity at high energies.
We have evaluated the new resonance scale by utilizing violation of perturbative unitarity, where we apply the fitting function used in the elastic pion-pion scattering to the phase shift in the present process.
We have also discussed the possibility that the phase shift can be measured by the asymmetry $A_\pm$ at the LHC and future hadron colliders.
The result is independent of the detail of composite models.

On the other hand, if we assume the "obsereved" composite model as one of the MCHMs, we need to go another approach to narrow down to a class of more specific models by experiments.
We have discussed the deviation patterns of the Higgs boson coupling constants as one of the ways to classify the types of the MCHMs.
We have also made a comprehensive list of the deviation patterns in a wide class of MCHMs.
We have found that the detailed study by using the deviation pattern can be an important alternative approach for proving the question whether the Higgs boson is a composite state or not.
Furthermore, it can be essentially important to distinguish a specific MCHM from the other new physics models.

We conclude that these two complementary approaches are very useful to explore the Higgs boson compositeness in future collider experiments.

\begin{acknowledgments}
This work was supported in part by Grant-in-Aid for Scientiﬁc Research from Japan Society for the Promotion of Science (JSPS), Nos. 22244031 (S.K.) and 24340046 (S.K. and T.S.), and from Ministry of Education, Culture, Sports, Science and Technology (MEXT), Japan, Nos. 23104006 (S.K.) and 23104011 (T.S.).
The work of N.M. was supported in part by the Sasakawa Scientific Research Grant from the Japan Science Society.
\end{acknowledgments}

\appendix


\section{Kinematics}
\label{sec:Kinematics}
We give the explicit kinematics of the decay product, which is sketched in Ref.~\cite{Murayama:2014yja}.
The process considered here is $WZ$ pair production by $u\bar d$ scattering, and they decay purely leptonic: $u(p_u)\bar d(p_d)\to W^+(p_W) Z(p_Z)\to \nu_l(p_\nu) l^+(p_{l_1}) l^-(p_{l_2}) l^+(p_{l_3})$. 
Regarding $u$ and $\bar d$ as massless, they only appear as left-handed state in this process.
Here we assign their momenta as follows:
\begin{eqnarray}
	u:      && p_u = \sqrt{S}/2~(1, -\sin\Theta, 0, \cos\Theta),\\
	\bar d: && p_d = \sqrt{S}/2~(1, \sin\Theta, 0, -\cos\Theta),\\
	W^+:	&& p_W = (E_W, 0, 0, p_V),\\
	Z:		&& p_Z = (E_Z, 0, 0, -p_V),\\
	\nu_l:	&& p_\nu = \sqrt{E_W^2-q_V^2}/2~(1, \sin\theta_1\cos\phi_1, \sin\theta_1\sin\phi_1, \cos\theta_1),\\
	l^+:	&& p_{l_1} = \sqrt{E_W^2-q_V^2}/2~(1, -\sin\theta_1\cos\phi_1, -\sin\theta_1\sin\phi_1, -\cos\theta_1),\\
	l^-:	&& p_{l_2} = \sqrt{E_Z^2-q_V^2}/2~(1, \sin\theta_2\cos\phi_2, \sin\theta_2\sin\phi_2, \cos\theta_2),\\
	l^+:	&& p_{l_3} = \sqrt{E_Z^2-q_V^2}/2~(1, -\sin\theta_2\cos\phi_2, -\sin\theta_2\sin\phi_2, -\cos\theta_2),
\end{eqnarray}
where we define $z$-axis along to the $W$-boson momentum direction, and $\Theta$ is the angle between $\vec p_u$ and $\vec p_W$.
The phase space of the final state leptons depends on two polar decay angles $(\theta_1,\theta_2)$ and two azimuthal decay angles $(\phi_1,\phi_2)$ from the production plane defined by $\hat n\sim \vec p_u\times \vec p_W$.
Therefore, $A_\pm$ defined in Eq.~(\ref{eq:asymmetry}) represents the asymmetry between the events that the charged lepton goes to "above" or "below" the production plane.

\section{Variations of the MCHMs}
\label{sec:models}
Here we list the matter sector of the effective Lagrangian and the Higgs potential in 
the models which we considered in this paper (see also Ref.~\cite{Carena:2014ria} excepting for MCHM$_{14}$).
In the MCHMs, the breaking pattern of the global symmetry is fixed as 
$SO(5)\times U(1)_X\to SO(4)\times U(1)_X$. 
Therefore the representations in which the SM fermions are embedded can be decomposed 
under $SO(4)\times U(1)_X\simeq SU(2)_L\times SU(2)_R\times U(1)_X$, 
as 
\begin{align}
	\pmb{5}_X\sim& (\pmb{2},\pmb{2})_X\oplus (\pmb{1},\pmb{1})_X\;,\nonumber\\
	\pmb{10}_X\sim& (\pmb{3},\pmb{1})_X\oplus (\pmb{1},\pmb{3})_X\oplus (\pmb{2},\pmb{2})_X\;,\\
	\pmb{14}_X\sim& (\pmb{3},\pmb{\bar{3}})_X\oplus (\pmb{2},\pmb{2})_X\oplus (\pmb{1},\pmb{1})_X\;,\nonumber
\end{align}
where the $X$ in the subscript denotes the charge for $U(1)_X$.

\subsection{$\text{MCHM}_5$}
All the quark fields are embedded into 5-representation.
We focus on the third generation quarks in the following. 
The quantum charges for $t_{L,R}$, and $b_{L,R}$ under $SU(2)_L\times SU(2)_R\times U(1)_X$ are assigned as
$t_L\sim (1/2,-1/2)_{2/3}$, $b_L\sim (-1/2, -1/2)_{2/3}$, 
$t_R\sim (0,0)_{2/3}$, and 
$b_R\sim (0,0)_{-1/3}$.
In the bracket, we write the quantum numbers corresponding to  $(SU(2)_L, SU(2)_R)_{U(1)_X}$.
The matter sector of the effective Lagrangian is 
\begin{align}
	\mathcal{L}_{\text{eff}}^{\text{matter}}=&
	\sum_{r=t_L,t_R,b_L,b_R}\overline{\Psi}_r^{(5)}\left[\slashed{p}\Pi_0^r+\Sigma^{\dagger}\slashed{p}\Pi_1^r\Sigma\right]\Psi_r^{(5)}
	\nonumber\\
	&
	+\overline{\Psi}_{t_L}^{(5)}\left[M_0^t+\Sigma^{\dagger}M_1^t\Sigma\right]\Psi_{t_R}^{(5)}
	+\overline{\Psi}_{b_L}^{(5)}\left[M_0^b+\Sigma^{\dagger}M_1^b\Sigma\right]\Psi_{b_R}^{(5)}+\text{h.c.}
	\;.
\end{align}
The effective Higgs potential takes the form as
\begin{equation}
	V_h\simeq \alpha \cos^2(h/f)+\beta\cos^2(h/f)\sin^2(h/f)\;.
	\label{eq:Vh5}
\end{equation}

\subsection{$\text{MCHM}_{10}$}
All the quark fields are embedded into 10-representation. 
The quantum charges for $t_{L,R}$, and $b_{L,R}$ under $SU(2)_L\times SU(2)_R\times U(1)_X$ are 
assigned as 
$t_L\sim (1/2,-1/2)_{2/3}$, $b_L\sim (-1/2, -1/2)_{2/3}$, 
$t_R\sim (0,0)_{2/3}$, and 
$b_R\sim (0,-1)_{2/3}$.
The matter sector of the effective Lagrangian is 
\begin{align}
	\mathcal{L}_{\text{eff}}^{\text{matter}}=&
	\sum_{r=q_L,t_R,b_R}\left[\overline{\Psi}_r^{(10)}\slashed{p}\Pi_0^r\Psi_r^{(10)}+(\Sigma\overline{\Psi}_r^{(10)})\slashed{p}\Pi_1^r(\Psi_r^{(10)}\Sigma^{\dagger})\right]
	\nonumber\\
	&
	+
	\overline{\Psi}_{q_L}^{(10)}M_0^t\Psi_{t_R}^{(10)}+(\Sigma\overline{\Psi}_{q_L}^{(10)})M_1^t(\Psi_{t_R}^{(10)}\Sigma^{\dagger})\nonumber\\
	&+\overline{\Psi}_{q_L}^{(10)}M_0^b\Psi_{b_R}^{(10)}+(\Sigma\overline{\Psi}_{q_L}^{(10)})M_1^b(\Psi_{b_R}^{(10)}\Sigma^{\dagger})
	+\text{h.c.}\;.
\end{align}
The effective Higgs potential takes the same form as one given in Eq.~(\ref{eq:Vh5}).

\subsection{$\text{MCHM}_{14}$}
All the quark fields are embedded into 14-representation. 
The quantum charges for $t_{L,R}$, and $b_{L,R}$ under $SU(2)_L\times SU(2)_R\times U(1)_X$ are 
assigned as 
$t_L\sim (1/2,-1/2)_{2/3}$, $b_L\sim (-1/2, -1/2)_{2/3}$, 
$t_R\sim (0,0)_{2/3}$, and 
$b_R\sim (0,-1)_{2/3}$.
The matter sector of the effective Lagrangian is 
\begin{align}
	\mathcal{L}_{\text{eff}}^{\text{matter}}=&
	\sum_{r=q_L,t_R,b_R}\left[\overline{\Psi}_r^{(14)}\slashed{p}\Pi_0^r\Psi_r^{(14)}
		+(\Sigma\overline{\Psi}_r^{(14)})\slashed{p}\Pi_1^r(\Psi_r^{(14)}\Sigma^{\dagger})
	+(\Sigma\overline{\Psi}_r^{(14)}\Sigma^{\dagger})\slashed{p}\Pi_2^r(\Sigma\Psi_r^{(14)}\Sigma^{\dagger})\right]
	\nonumber\\
	&
	+\overline{\Psi}_{q_L}^{(14)}M_0^t\Psi_{t_R}^{(14)}
	+(\Sigma\overline{\Psi}_{q_L}^{(14)})M_1^t(\Psi_{t_R}^{(14)}\Sigma^{\dagger})
	+(\Sigma\overline{\Psi}_{q_L}^{(14)}\Sigma^{\dagger})M_2^t(\Sigma\Psi_{t_R}^{(14)}\Sigma^{\dagger})\nonumber\\
	&
	+\overline{\Psi}_{q_L}^{(14)}M_0^b\Psi_{b_R}^{(14)}
	+(\Sigma\overline{\Psi}_{q_L}^{(14)})M_1^b(\Psi_{b_R}^{(14)}\Sigma^{\dagger})
	+(\Sigma\overline{\Psi}_{q_L}^{(14)}\Sigma^{\dagger})M_2^b(\Sigma\Psi_{b_R}^{(14)}\Sigma^{\dagger})
	+\text{h.c.}\;.
\end{align}
The effective Higgs potential takes the form as 
\begin{equation}
	V_h\simeq \alpha \sin^2(h/f)+\beta\sin^4(h/f)+\gamma\sin^6(h/f)\;.
	\label{eq:Vh14}
\end{equation}

\subsection{$\text{MCHM}_\text{5-1-10}$}
The quantum charges for $t_{L,R}$, and $b_{L,R}$ under $SU(2)_L\times SU(2)_R\times U(1)_X$ are 
assigned as 
$t_L\sim (1/2,-1/2)_{2/3}$, $b_L\sim (-1/2, -1/2)_{2/3}$, 
$t_R\sim (0,0)_{2/3}$, and 
$b_R\sim (0,-1)_{2/3}$.
The matter sector of the effective Lagrangian is 
\begin{align}
	\mathcal{L}_{\text{eff}}^{\text{matter}}=&
	\overline{\Psi}_{q_L}^{(5)}\slashed{p}\Pi_0^{q_L}\Psi_{q_L}^{(5)}
	+(\overline{\Psi}_{q_L}^{(5)}\Sigma^{\dagger})\slashed{p}\Pi_1^{q_L}(\Sigma\Psi_{q_L}^{(5)})
	\nonumber\\
	&
	+\overline{\Psi}_{t_R}^{(1)}\slashed{p}\Pi_0^{t_R}\Psi_{t_R}^{(1)}
	\nonumber\\
	&
	+\overline{\Psi}_{b_R}^{(10)}\slashed{p}\Pi_0^{b_R}\Psi_{b_R}^{(10)}
	+(\Sigma\overline{\Psi}_{b_R}^{(10)})\slashed{p}\Pi_1^{b_R}(\Psi_{b_R}^{(10)}\Sigma^{\dagger})\nonumber\\
	&
	+(\overline{\Psi}_{q_L}^{(5)}\Sigma^{\dagger})M_1^t\Psi_{t_R}^{(1)}
	+\overline{\Psi}_{q_L}^{(5)}M_1^b(\Psi_{b_R}^{(10)}\Sigma^{\dagger})
	+\text{h.c.}\;.
\end{align}
The effective Higgs potential takes the form as 
\begin{equation}
	V_h\simeq -\beta \sin^2(h/f)\;.
\end{equation}
However, the electroweak symmetry breaking cannot occur with this potential and 
this model is not a realistic model.

\subsection{$\text{MCHM}_\text{5-5-10}$}
The quantum charges for $t_{L,R}$, and $b_{L,R}$ under $SU(2)_L\times SU(2)_R\times U(1)_X$ are 
assigned as 
$t_L\sim (1/2,-1/2)_{2/3}$, $b_L\sim (-1/2, -1/2)_{2/3}$, 
$t_R\sim (0,0)_{2/3}$, and 
$b_R\sim (0,-1)_{2/3}$.
The matter sector of the effective Lagrangian is 
\begin{align}
	\mathcal{L}_{\text{eff}}^{\text{matter}}=&
	\overline{\Psi}_{q_L}^{(5)}\slashed{p}\Pi_0^{q_L}\Psi_{q_L}^{(5)}
	+(\overline{\Psi}_{q_L}^{(5)}\Sigma^{\dagger})\slashed{p}\Pi_1^{q_L}(\Sigma\Psi_{q_L}^{(5)})
	\nonumber\\
	&
	+\overline{\Psi}_{t_R}^{(5)}\slashed{p}\Pi_0^{t_R}\Psi_{t_R}^{(5)}
	+(\overline{\Psi}_{t_R}^{(5)}\Sigma^{\dagger})\slashed{p}\Pi_1^{t_R}(\Sigma\Psi_{t_R}^{(5)})\nonumber\\
	&+\overline{\Psi}_{b_R}^{(10)}\slashed{p}\Pi_0^{b_R}\Psi_{b_R}^{(10)}
	+(\Sigma\overline{\Psi}_{b_R}^{(10)})\slashed{p}\Pi_1^{b_R}(\Psi_{b_R}^{(10)}\Sigma^{\dagger})\nonumber\\
	&+\overline{\Psi}_{q_L}^{(5)}M_0^t\Psi_{t_R}^{(5)}
	+(\overline{\Psi}_{q_L}^{(5)}\Sigma^{\dagger})M_1^t(\Sigma\Psi_{t_R}^{(5)})
	+\overline{\Psi}_{q_L}^{(5)}M_1^b(\Psi_{b_R}^{(10)}\Sigma^{\dagger})
	+\text{h.c.}\;.
\end{align}
The effective Higgs potential takes the same form as one given in Eq.~(\ref{eq:Vh5}).

\subsection{$\text{MCHM}_\text{5-10-10}$}
The quantum charges for $t_{L,R}$, and $b_{L,R}$ under $SU(2)_L\times SU(2)_R\times U(1)_X$ are 
assigned as 
$t_L\sim (1/2,-1/2)_{2/3}$, $b_L\sim (-1/2, -1/2)_{2/3}$, 
$t_R\sim (0,0)_{2/3}$, and 
$b_R\sim (0,-1)_{2/3}$.
The matter sector of the effective Lagrangian is 
\begin{align}
	\mathcal{L}_{\text{eff}}^{\text{matter}}=&
	\overline{\Psi}_{q_L}^{(5)}\slashed{p}\Pi_0^{q_L}\Psi_{q_L}^{(5)}
	+(\overline{\Psi}_{q_L}^{(5)}\Sigma^{\dagger})\slashed{p}\Pi_1^{q_L}(\Sigma\Psi_{q_L}^{(5)})
	\nonumber\\
	&
	+\overline{\Psi}_{t_R}^{(10)}\slashed{p}\Pi_0^{t_R}\Psi_{t_R}^{(10)}
	+(\Sigma\overline{\Psi}_{t_R}^{(10)})\slashed{p}\Pi_1^{t_R}(\Psi_{t_R}^{(10)}\Sigma^{\dagger})\nonumber\\
	&+\overline{\Psi}_{b_R}^{(10)}\slashed{p}\Pi_0^{b_R}\Psi_{b_R}^{(10)}
	+(\Sigma\overline{\Psi}_{b_R}^{(10)})\slashed{p}\Pi_1^{b_R}(\Psi_{b_R}^{(10)}\Sigma^{\dagger})\nonumber\\
	&
	+\overline{\Psi}_{q_L}^{(5)}M_1^t(\Psi_{t_R}^{(10)}\Sigma^{\dagger})
	+\overline{\Psi}_{q_L}^{(5)}M_1^b(\Psi_{b_R}^{(10)}\Sigma^{\dagger})
	+\text{h.c.}\;.
\end{align}
The effective Higgs potential takes the same form as one given in Eq.~(\ref{eq:Vh5}).

\subsection{$\text{MCHM}_\text{5-14-10}$}
The quantum charges for $t_{L,R}$, and $b_{L,R}$ under $SU(2)_L\times SU(2)_R\times U(1)_X$ are 
assigned as 
$t_L\sim (1/2,-1/2)_{2/3}$, $b_L\sim (-1/2, -1/2)_{2/3}$, 
$t_R\sim (0,0)_{2/3}$, and 
$b_R\sim (0,-1)_{2/3}$.
The matter sector of the effective Lagrangian is 
\begin{align}
	\mathcal{L}_{\text{eff}}^{\text{matter}}=&
	\overline{\Psi}_{q_L}^{(5)}\slashed{p}\Pi_0^{q_L}\Psi_{q_L}^{(5)}
	+(\overline{\Psi}_{q_L}^{(5)}\Sigma^{\dagger})\slashed{p}\Pi_1^{q_L}(\Sigma\Psi_{q_L}^{(5)})
	\nonumber\\
	&
	+\overline{\Psi}_{t_R}^{(14)}\slashed{p}\Pi_0^{t_R}\Psi_{t_R}^{(14)}
	+(\Sigma\overline{\Psi}_{t_R}^{(14)})\slashed{p}\Pi_1^{t_R}(\Psi_{t_R}^{(14)}\Sigma^{\dagger})
	+(\Sigma\overline{\Psi}_{t_R}^{(14)}\Sigma^{\dagger})\slashed{p}\Pi_2^{t_R}(\Sigma\Psi_{t_R}^{(14)}\Sigma^{\dagger})
	\nonumber\\
	&
	+\overline{\Psi}_{b_R}^{(10)}\slashed{p}\Pi_0^{b_R}\Psi_{b_R}^{(10)}
	+(\Sigma\overline{\Psi}_{b_R}^{(10)})\slashed{p}\Pi_1^{b_R}(\Psi_{b_R}^{(10)}\Sigma^{\dagger})\nonumber\\
	&
	+\overline{\Psi}_{q_L}^{(5)}M_1^t(\Psi_{t_R}^{(14)}\Sigma^{\dagger})
	+(\overline{\Psi}_{q_L}^{(5)}\Sigma^{\dagger})M_2^t(\Sigma\Psi_{t_R}^{(14)}\Sigma^{\dagger})
	+\overline{\Psi}_{q_L}^{(5)}M_1^b(\Psi_{b_R}^{(10)}\Sigma^{\dagger})
	+\text{h.c.}\;.
\end{align}
The effective Higgs potential takes the same form as one given in Eq.~(\ref{eq:Vh14}).

\subsection{$\text{MCHM}_\text{10-5-10}$}
The quantum charges for $t_{L,R}$, and $b_{L,R}$ under $SU(2)_L\times SU(2)_R\times U(1)_X$ are 
assigned as 
$t_L\sim (1/2,-1/2)_{2/3}$, $b_L\sim (-1/2, -1/2)_{2/3}$, 
$t_R\sim (0,0)_{2/3}$, and 
$b_R\sim (0,-1)_{2/3}$.
The matter sector of the effective Lagrangian is 
\begin{align}
	\mathcal{L}_{\text{eff}}^{\text{matter}}=&
	\overline{\Psi}_{q_L}^{(10)}\slashed{p}\Pi_0^{q_L}\Psi_{q_L}^{(10)}
	+(\Sigma\overline{\Psi}_{q_L}^{(10)})\slashed{p}\Pi_1^{q_L}(\Psi_{q_L}^{(10)}\Sigma^{\dagger})
	\nonumber\\
	&
	+\overline{\Psi}_{t_R}^{(5)}\slashed{p}\Pi_0^{t_R}\Psi_{t_R}^{(5)}
	+(\overline{\Psi}_{t_R}^{(5)}\Sigma^{\dagger})\slashed{p}\Pi_1^{t_R}(\Sigma\Psi_{t_R}^{(5)})
	\nonumber\\
	&
	+\overline{\Psi}_{b_R}^{(10)}\slashed{p}\Pi_0^{b_R}\Psi_{b_R}^{(10)}
	+(\Sigma\overline{\Psi}_{b_R}^{(10)})\slashed{p}\Pi_1^{b_R}(\Psi_{b_R}^{(10)}\Sigma^{\dagger})\nonumber\\
	&
	+(\Sigma\overline{\Psi}_{q_L}^{(10)})M_1^t\Psi_{t_R}^{(5)}
	+\overline{\Psi}_{q_L}^{(10)}M_0^b\Psi_{b_R}^{(10)}
	+(\Sigma\overline{\Psi}_{q_L}^{(10)})M_1^b(\Psi_{b_R}^{(10)}\Sigma^{\dagger})
	+\text{h.c.}\;.
\end{align}
The effective Higgs potential takes the same form as one given in Eq.~(\ref{eq:Vh5}).

\subsection{$\text{MCHM}_\text{10-14-10}$}
The quantum charges for $t_{L,R}$, and $b_{L,R}$ under $SU(2)_L\times SU(2)_R\times U(1)_X$ are 
assigned as 
$t_L\sim (1/2,-1/2)_{2/3}$, $b_L\sim (-1/2, -1/2)_{2/3}$, 
$t_R\sim (0,0)_{2/3}$, and 
$b_R\sim (0,-1)_{2/3}$.
The matter sector of the effective Lagrangian is 
\begin{align}
	\mathcal{L}_{\text{eff}}^{\text{matter}}=&
	\overline{\Psi}_{q_L}^{(10)}\slashed{p}\Pi_0^{q_L}\Psi_{q_L}^{(10)}
	+(\Sigma\overline{\Psi}_{q_L}^{(10)})\slashed{p}\Pi_1^{q_L}(\Psi_{q_L}^{(10)}\Sigma^{\dagger})\nonumber\\
	&
	+\overline{\Psi}_{t_R}^{(14)}\slashed{p}\Pi_0^{t_R}\Psi_{t_R}^{(14)}
	+(\Sigma\overline{\Psi}_{t_R}^{(14)})\slashed{p}\Pi_1^{t_R}(\Psi_{t_R}^{(14)}\Sigma^{\dagger})
	+(\Sigma\overline{\Psi}_{t_R}^{(14)}\Sigma^{\dagger})\slashed{p}\Pi_2^{t_R}(\Sigma\Psi_{t_R}^{(14)}\Sigma^{\dagger})
	\nonumber\\
	&
	+\overline{\Psi}_{b_R}^{(10)}\slashed{p}\Pi_0^{b_R}\Psi_{b_R}^{(10)}
	+(\Sigma\overline{\Psi}_{b_R}^{(10)})\slashed{p}\Pi_1^{b_R}(\Psi_{b_R}^{(10)}\Sigma^{\dagger})\nonumber\\
	&
	+(\Sigma\overline{\Psi}_{q_L}^{(10)})M_1^t(\Psi_{t_R}^{(14)}\Sigma^{\dagger})
	+(\Sigma\overline{\Psi}_{q_L}^{(10)})M_1^b(\Psi_{b_R}^{(10)}\Sigma^{\dagger})
	+\text{h.c.}\;.
\end{align}
The effective Higgs potential takes the same form as one given in Eq.~(\ref{eq:Vh14}).

\subsection{$\text{MCHM}_\text{14-1-10}$}
The quantum charges for $t_{L,R}$, and $b_{L,R}$ under $SU(2)_L\times SU(2)_R\times U(1)_X$ are 
assigned as 
$t_L\sim (1/2,-1/2)_{2/3}$, $b_L\sim (-1/2, -1/2)_{2/3}$, 
$t_R\sim (0,0)_{2/3}$, and 
$b_R\sim (0,-1)_{2/3}$.
The matter sector of the effective Lagrangian is 
\begin{align}
	\mathcal{L}_{\text{eff}}^{\text{matter}}=&
	\overline{\Psi}_{q_L}^{(14)}\slashed{p}\Pi_0^{q_L}\Psi_{q_L}^{(14)}
	+(\Sigma\overline{\Psi}_{q_L}^{(14)})\slashed{p}\Pi_1^{q_L}(\Psi_{q_L}^{(14)}\Sigma^{\dagger})
	+(\Sigma\overline{\Psi}_{q_L}^{(14)}\Sigma^{\dagger})\slashed{p}\Pi_2^{q_L}(\Sigma\Psi_{q_L}^{(14)}\Sigma^{\dagger})
	\nonumber\\
	&+\overline{\Psi}_{t_R}^{(1)}\slashed{p}\Pi_0^{t_R}\Psi_{t_R}^{(1)}
	\nonumber\\
	&
	+\overline{\Psi}_{b_R}^{(10)}\slashed{p}\Pi_0^{b_R}\Psi_{b_R}^{(10)}
	+(\Sigma\overline{\Psi}_{b_R}^{(10)})\slashed{p}\Pi_1^{b_R}(\Psi_{b_R}^{(10)}\Sigma^{\dagger})\nonumber\\
	&
	+(\Sigma\overline{\Psi}_{q_L}^{(14)}\Sigma^{\dagger})M_2^t\Psi_{t_R}^{(1)}
	+(\Sigma\overline{\Psi}_{q_L}^{(14)})M_1^b(\Psi_{b_R}^{(10)}\Sigma^{\dagger})
	+\text{h.c.}\;.
\end{align}
The effective Higgs potential takes the same form as one given in Eq.~(\ref{eq:Vh5}).
\subsection{$\text{MCHM}_\text{14-5-10}$}
The quantum charges for $t_{L,R}$, and $b_{L,R}$ under $SU(2)_L\times SU(2)_R\times U(1)_X$ are 
assigned as 
$t_L\sim (1/2,-1/2)_{2/3}$, $b_L\sim (-1/2, -1/2)_{2/3}$, 
$t_R\sim (0,0)_{2/3}$, and 
$b_R\sim (0,-1)_{2/3}$.
The matter sector of the effective Lagrangian is 
\begin{align}
	\mathcal{L}_{\text{eff}}^{\text{matter}}=&
	\overline{\Psi}_{q_L}^{(14)}\slashed{p}\Pi_0^{q_L}\Psi_{q_L}^{(14)}
	+(\Sigma\overline{\Psi}_{q_L}^{(14)})\slashed{p}\Pi_1^{q_L}(\Psi_{q_L}^{(14)}\Sigma^{\dagger})
	+(\Sigma\overline{\Psi}_{q_L}^{(14)}\Sigma^{\dagger})\slashed{p}\Pi_2^{q_L}(\Sigma\Psi_{q_L}^{(14)}\Sigma^{\dagger})
	\nonumber\\
	&
	+\overline{\Psi}_{t_R}^{(5)}\slashed{p}\Pi_0^{t_R}\Psi_{t_R}^{(5)}
	+(\overline{\Psi}_{t_R}^{(5)}\Sigma^{\dagger})\slashed{p}\Pi_1^{t_R}(\Sigma\Psi_{t_R}^{(5)})\nonumber\\
	&
	+\overline{\Psi}_{b_R}^{(10)}\slashed{p}\Pi_0^{b_R}\Psi_{b_R}^{(10)}
	+(\Sigma\overline{\Psi}_{b_R}^{(10)})\slashed{p}\Pi_1^{b_R}(\Psi_{b_R}^{(10)}\Sigma^{\dagger})\nonumber\\
	&
	+(\Sigma\overline{\Psi}_{q_L}^{(14)})M_1^t\Psi_{t_R}^{(5)}
	+(\Sigma\overline{\Psi}_{q_L}^{(14)}\Sigma^{\dagger})M_2^t(\Sigma\Psi_{t_R}^{(5)})
	+(\Sigma\overline{\Psi}_{q_L}^{(14)})M_1^b(\Psi_{b_R}^{(10)}\Sigma^{\dagger})
	+\text{h.c.}\;.
\end{align}
The effective Higgs potential takes the same form as one given in Eq.~(\ref{eq:Vh14}).

\subsection{$\text{MCHM}_\text{14-10-10}$}
The quantum charges for $t_{L,R}$, and $b_{L,R}$ under $SU(2)_L\times SU(2)_R\times U(1)_X$ are 
assigned as 
$t_L\sim (1/2,-1/2)_{2/3}$, $b_L\sim (-1/2, -1/2)_{2/3}$, 
$t_R\sim (0,0)_{2/3}$, and 
$b_R\sim (0,-1)_{2/3}$.
The matter sector of the effective Lagrangian is 
\begin{align}
	\mathcal{L}_{\text{eff}}^{\text{matter}}=&
	\overline{\Psi}_{q_L}^{(14)}\slashed{p}\Pi_0^{q_L}\Psi_{q_L}^{(14)}
	+(\Sigma\overline{\Psi}_{q_L}^{(14)})\slashed{p}\Pi_1^{q_L}(\Psi_{q_L}^{(14)}\Sigma^{\dagger})
	+(\Sigma\overline{\Psi}_{q_L}^{(14)}\Sigma^{\dagger})\slashed{p}\Pi_2^{q_L}(\Sigma\Psi_{q_L}^{(14)}\Sigma^{\dagger})
	\nonumber\\
	&
	+\overline{\Psi}_{t_R}^{(10)}\slashed{p}\Pi_0^{t_R}\Psi_{t_R}^{(10)}
	+(\Sigma\overline{\Psi}_{t_R}^{(10)})\slashed{p}\Pi_1^{t_R}(\Psi_{t_R}^{(10)}\Sigma^{\dagger})\nonumber\\
	&
	+\overline{\Psi}_{b_R}^{(10)}\slashed{p}\Pi_0^{b_R}\Psi_{b_R}^{(10)}
	+(\Sigma\overline{\Psi}_{b_R}^{(10)})\slashed{p}\Pi_1^{b_R}(\Psi_{b_R}^{(10)}\Sigma^{\dagger})\nonumber\\
	&
	+(\Sigma\overline{\Psi}_{q_L}^{(14)})M_1^t(\Psi_{t_R}^{(10)}\Sigma^{\dagger})
	+(\Sigma\overline{\Psi}_{q_L}^{(14)})M_1^b(\Psi_{b_R}^{(10)}\Sigma^{\dagger})
	+\text{h.c.}\;.
\end{align}
The effective Higgs potential takes the same form as one given in Eq.~(\ref{eq:Vh14}).

\subsection{$\text{MCHM}_\text{14-14-10}$}
The quantum charges for $t_{L,R}$, and $b_{L,R}$ under $SU(2)_L\times SU(2)_R\times U(1)_X$ are 
assigned as 
$t_L\sim (1/2,-1/2)_{2/3}$, $b_L\sim (-1/2, -1/2)_{2/3}$, 
$t_R\sim (0,0)_{2/3}$, and 
$b_R\sim (0,-1)_{2/3}$.
The matter sector of the effective Lagrangian is 
\begin{align}
	\mathcal{L}_{\text{eff}}^{\text{matter}}=&
	\overline{\Psi}_{q_L}^{(14)}\slashed{p}\Pi_0^{q_L}\Psi_{q_L}^{(14)}
	+(\Sigma\overline{\Psi}_{q_L}^{(14)})\slashed{p}\Pi_1^{q_L}(\Psi_{q_L}^{(14)}\Sigma^{\dagger})
	+(\Sigma\overline{\Psi}_{q_L}^{(14)}\Sigma^{\dagger})\slashed{p}\Pi_2^{q_L}(\Sigma\Psi_{q_L}^{(14)}\Sigma^{\dagger})
	\nonumber\\
	&
	+\overline{\Psi}_{t_R}^{(14)}\slashed{p}\Pi_0^{t_R}\Psi_{t_R}^{(14)}
	+(\Sigma\overline{\Psi}_{t_R}^{(14)})\slashed{p}\Pi_1^{t_R}(\Psi_{t_R}^{(14)}\Sigma^{\dagger})
	+(\Sigma\overline{\Psi}_{t_R}^{(14)}\Sigma^{\dagger})\slashed{p}\Pi_2^{t_R}(\Sigma\Psi_{t_R}^{(14)}\Sigma^{\dagger})
	\nonumber\\
	&
	+\overline{\Psi}_{b_R}^{(10)}\slashed{p}\Pi_0^{b_R}\Psi_{b_R}^{(10)}
	+(\Sigma\overline{\Psi}_{b_R}^{(10)})\slashed{p}\Pi_1^{b_R}(\Psi_{b_R}^{(10)}\Sigma^{\dagger})\nonumber\\
	&
	+\overline{\Psi}_{q_L}^{(14)}M_0^t\Psi_{t_R}^{(14)}
	+(\Sigma\overline{\Psi}_{q_L}^{(14)})M_1^t(\Psi_{t_R}^{(14)}\Sigma^{\dagger})
	+(\Sigma\overline{\Psi}_{q_L}^{(14)}\Sigma^{\dagger})M_2^t(\Sigma\Psi_{t_R}^{(14)}\Sigma^{\dagger})
	\nonumber\\
	&
	+(\Sigma\overline{\Psi}_{q_L}^{(14)})M_1^b(\Psi_{b_R}^{(10)}\Sigma^{\dagger})
	+\text{h.c.}\;.
\end{align}
The effective Higgs potential takes the same form as one given in Eq.~(\ref{eq:Vh14}).

\bibliography{Reference}

\end{document}